\documentclass[11pt]{amsart}
\usepackage[dvips]{graphics}
\DeclareGraphicsExtensions{.eps.gz,.eps,.epsi.gz,.epsi,.ps,.ps.gz}
\usepackage{epsfig}
\usepackage{color}
\graphicspath{{fig/}} \textwidth=6.5in \textheight=9in

\usepackage{latexsym}
\usepackage{graphicx}
\usepackage{amsmath}
\usepackage{amsfonts}
\usepackage{amssymb}
\usepackage{mathrsfs}
\usepackage[lofdepth,lotdepth]{subfig}
\usepackage{multirow}
\usepackage{booktabs}
\usepackage{lscape}

\usepackage{fancyhdr}
\usepackage{amsmath}
\usepackage{latexsym}
\usepackage{mathrsfs} 
\usepackage{amsfonts}
\usepackage{amssymb}

\usepackage{graphicx}
\usepackage{verbatim}
\usepackage{geometry}

\newcommand{\bel}{\begin{eqnarray}\label}
\newcommand{\eel}{\end{eqnarray}}
\newcommand{\bes}{\begin{eqnarray*}}
\newcommand{\ees}{\end{eqnarray*}}

\def\benu{\begin{enumerate}}
\def\eenu{\end{enumerate}}

\def\argmin{\mathop{\rm arg\, min}}

\def\real{{\mathbb{R}}}

\def\complex{\mathop{{\rm I}\kern-.58em\hbox{\rm C}}\nolimits}
\def\pa{\partial}

\def\diag{\hbox{diag}}

\def\rank{\hbox{rank}}

\def\sgn{\hbox{sgn}}\def\sgn{\hbox{\rm sgn}}

\def\Cov{\hbox{Cov}}

\def\supp{\hbox{supp}}

\def\mathbold{\boldsymbol} 

\def\ba{\mathbold{a}}

\def\scrA{{\mathscr A}}

\def\bb{\mathbold{b}}

\def\scrB{{\mathscr B}}

\def\scrC{{\mathscr C}}

\def\scrC{{\mathscr C}}

\def\bh{\mathbold{h}}

\def\bI{\mathbold{I}}

\def\ptil{\widetilde{p}}

\def\bP{\mathbold{P}}

\def\bQ{\mathbold{Q}}

\def\bs{\mathbold{s}}

\def\Shat{\widehat{S}}

\def\that{\widehat{t}}\def\ttil{\widetilde{t}}

\def\bu{\mathbold{u}}

\def\bv{\mathbold{v}}

\def\bV{\mathbold{V}}

\def\bw{\mathbold{w}}

\def\bx{\mathbold{x}}\def\xtil{\widetilde{x}}
\def\tbx{{\widetilde{\bx}}}
\def\bX{\mathbold{X}}\def\tbX{{\widetilde{\bX}}}
\def\scrX{{\mathscr X}}\def\Xtil{{\widetilde X}}

\def\by{\mathbold{y}}

\def\bz{\mathbold{z}}\def\zhat{\widehat{z}}

\def\bbeta{\mathbold{\beta}}\def\hbeta{\widehat{\beta}}
\def\tbeta{\widetilde{\beta}}
\def\hbbeta{{\widehat{\bbeta}}}

\def\bgamma{\mathbold{\gamma}}
\def\tgamma{\widetilde{\gamma}}
\def\hbgamma{{\widehat{\bgamma}}}

\def\bdelta{\mathbold{\delta}}

\def\ep{\varepsilon}\def\eps{\epsilon}\def\veps{\varepsilon}
\def\bep{\mathbold{\ep}}

\def\lam{\lambda}

\def\hsigma{\widehat{\sigma}}
\def\tsigma{\widetilde{\sigma}}

\def\bSigma{\mathbold{\Sigma}}\def\hbSigma{{\widehat{\bSigma}}}

\newtheorem{theorem}{Theorem}
\newtheorem{lemma}{Lemma}
\newtheorem{proposition}{Proposition}

\newtheorem{remark}{Remark}

\def\argmin{\mathop{\rm arg\, min}}
\def\real{\mathop{{\rm I}\kern-.2em\hbox{\rm R}}\nolimits}

\def\R{{\mathbb{R}}}

\def\supp{\hbox{supp}}

\def\tveps{{\widetilde \veps}}

\begin{document}

\title{Confidence Intervals for Low-Dimensional Parameters in High-Dimensional Linear Models} 
\author{Cun-Hui Zhang}
\address{Department of Statistics and Biostatistics, Hill Center, Busch Campus, 
Rutgers University, Piscataway, NJ 08854, USA}
\email{czhang@stat.rutgers.edu}
\author{Stephanie S. Zhang}\address{Department of Statistics, Columbia University, 
New York, NY 10027}\email{sszhang@stat.columbia.edu}

\begin{abstract}
The purpose of this paper is to propose methodologies for statistical inference of low-dimensional 
parameters with high-dimensional data.  We focus on constructing confidence intervals for 
individual coefficients and linear combinations of several of them in a linear regression model, although 
our ideas are applicable in a much broader context. The theoretical results presented here provide 
sufficient conditions for the asymptotic normality of the proposed estimators along with a
consistent estimator for their finite-dimensional covariance matrices. These sufficient conditions 
allow the number of variables to far exceed the sample size.  The simulation results 
presented here demonstrate the accuracy of the coverage probability of the proposed confidence 
intervals, strongly supporting the theoretical results.
\end{abstract}

\maketitle
\thispagestyle{empty}


\section{Introduction}
High-dimensional data is an intense area of research in statistics and machine learning, 
due to the rapid development of information technologies and their applications in scientific 
experiments and everyday life.  Numerous large, complex datasets have been collected and 
are waiting to be analyzed; meanwhile, an enormous effort has been mounted in order to meet 
this challenge by researchers and practitioners in statistics, computer science, and other disciplines. 
A great number of statistical methods, algorithms, and theories 
have been developed for the prediction and classification of future outcomes, 
the estimation of high-dimensional objects, 
and the selection of important variables or features for further scientific experiments and engineering applications. 
However, statistical inference with high-dimensional data is still 
largely untouched, due to the complexity of the sampling distributions of existing estimators. 
This is particularly the case in the context of the so called large-p-smaller-n problem, where 
the dimension of the data $p$ is greater than the sample size $n$. 

Regularized linear regression is one of the best understood statistical problems in 
high-dimensional data. Important work has been done in formulation of problems, 
development of methodologies and algorithms, and theoretical understanding 
of their performance under sparsity assumptions on the regression coefficients. 
This includes $\ell_1$ regularized methods 
\cite{Tibshirani96, ChenDS01,
GreenshteinR04,Greenshtein06,MeinshausenB06,Tropp06,ZhaoY06,
CandesT07,
ZhangH08,BickelRT09,Koltchinskii09,MeinshausenY09,vandeGeerB09,Wainwright09,Zhang09-l1,
YeZ10,KoltchinskiiLT11,SunZ11}, 
nonconvex penalized methods 
\cite{FrankF93,FanL01,FanP04,
KiChOh08,Zhang10-mc+,
ZhangZ11}, 
greedy methods \cite{
Zhang11-foba}, 
adaptive methods \cite{Zou06,HuangMZ08,ZouL08,
Zhang11-multistage,ZhangZ11}, 
screening methods \cite{FanL08}, and more. For further discussion, we refer to related sections in 
\cite{BuhlmannGeer11} and recent reviews in \cite{FanL10,ZhangZ11}. 

Among existing results, variable selection consistency is most relevant to statistical inference. 
An estimator is variable selection consistent if it selects the oracle model composed of 
exactly the set of variables with nonzero 
regression coefficients. In the large-p-smaller-n setting, 
variable selection consistency has been established under incoherence and other 
$\ell_\infty$-type conditions on the design matrix for the Lasso 
\cite{MeinshausenB06,Tropp06,ZhaoY06,Wainwright09}, 
and under sparse eigenvalue or $\ell_2$-type conditions for nonconvex methods 
\cite{FanP04,Zhang10-mc+,Zhang11-foba,Zhang11-multistage,ZhangZ11}. 
Another approach in variable selection with high-dimensional data involves subsampling or randomization,  
including notably the stability selection method proposed in \cite{MeinshausenB10}. 
Since the oracle model is typically 
assumed to be of smaller order 
in dimension than the sample size $n$ in selection consistency theory, consistent variable selection 
allows a great reduction of the complexity of the analysis from a large-p-smaller-n problem 
to one involving the oracle set of variables only. 
Consequently, taking the least squares estimator on the selected set of variables if necessary, statistical inference can be justified in the smaller oracle model.  

However, statistical inference based on selection consistency theory typically requires a uniform 
signal strength condition that all nonzero regression coefficients be greater in magnitude 
than an inflated noise level to take model uncertainty into account. 
This inflated noise level can be written as $C\sigma\sqrt{(2/n)\log p}$, 
where $\sigma$ is the noise level with each response. 
Based on the sharpest existing results, $C\ge 1/2$ is required for variable selection consistency with a 
general standardized design matrix \cite{Wainwright09b,Zhang10-mc+}.
This uniform signal strength condition is, unfortunately, seldom supported by either the data or the 
underlying science in applications when the presence of weak signals cannot be ruled out. 
Without this uniform signal strength assumption, consistent estimation of the distribution 
of the least squares estimator after model selection is impossible \cite{LeebP06}. 
Conservative statistical inference after model selection or classification has been considered 
in \cite{BerkBZ10,LaberM11}. 
However, such conservative methods 
may not yield sufficiently accurate confidence regions or p-values for common applications 
with a large number of variables.

We propose a low-dimensional projection (LDP) approach to constructing confidence intervals 
for regression coefficients without assuming the uniform signal strength condition. 
We provide theoretical justifications for the use of the proposed confidence interval for a 
preconceived regression coefficient or a contrast depending on a small number of regression coefficients.
We believe that in the presence of potentially many nonzero coefficients {of small or moderate magnitude},
construction of a confidence interval for such a preconceived parameter is an important 
problem in and of itself and was open before our paper \cite{LeebP06}, 
but the proposed method is not limited to this application.

Our theoretical work also justifies the use of LDP confidence intervals simultaneously with 
multiplicity adjustment. 
In the absence of a preconceived parameter of interest, the proposed simultaneous confidence 
intervals provide more information about the unknown regression coefficients than variable selection,
but this is not the main point. 

The most important difference between the proposed LDP and existing 
variable selection approaches concerns the requirement known as the uniform signal strength condition. 
As we have mentioned earlier, variable selection consistency requires all nonzero regression 
coefficients be greater than $C\sigma\sqrt{(2/n)\log p}$, with $C\ge 1/2$ at the least. 
This is a necessity for the simultaneous correct selection of {\it all zero or nonzero} coefficients. 
If this criterion is the goal, we can not do better than technical improvements over existing methods. 
However, a main complaint about the variable selection approach is the practicality of the
uniform signal strength condition, and the crucial difference between the two approaches 
is precisely in the case where the condition fails to hold. 
Without the condition, neither large nor zero coefficients are guaranteed to be 
correctly selected by existing variable selection methods in the presence of
potentially many nonzero coefficients below the radar screen, 
but the proposed method can. 
The power of the proposed method is small for testing small nonzero coefficients, 
but this is unavoidable {and does not affect the correct selection of other variables}.
In this sense, the proposed {confidence intervals} decompose the variable 
selection problem into multiple marginal testing problems for individual coefficients as Gaussian means. 

\section{Methodology} 
We develop methodologies and algorithms for the construction of confidence intervals 
for the individual regression coefficients and their linear combinations in the linear model 
\bel{LM}
\by = \bX\bbeta + \bep,\; \bep \sim \mathcal{N}(0, \sigma^2 \bI),
\eel
where $\by\in\R^n$ is a response vector, $\bX=(\bx_1,\ldots,\bx_p)\in\R^{n\times p}$ is a 
design matrix with columns $\bx_j$, and $\bbeta = (\beta_1,\ldots,\beta_p)^T$ is a 
vector of unknown regression coefficients. 
When $\rank(\bX) < p$, $\bbeta$ is unique under proper conditions on the sparsity of $\bbeta$ and 
regularity of $\bX$, but not in general. 
To simplify the discussion, we standardize the design to $\|\bx_j\|_2^2=n$. 
The design matrix $\bX$ is assumed to be deterministic throughout the paper, 
except in Subsection~\ref{subsec-random-design}.  

The following notation will be used. For real numbers $x$ and $y$, 
$x\wedge y = \min(x,y)$, $x\vee y =\max(x,y)$, $x_+=x\vee 0$, and $x_-=(-x)_+$. 
For vectors $\bv=(v_1,\ldots,v_m)$ of any dimension, $\supp(\bv)=\{j: v_j\neq 0\}$, 
$\|\bv\|_0=|\supp(\bv)|=\#\{j: v_j\neq 0\}$, and  
$\|\bv\|_q = \{\sum_j|v_j|^q)^{1/q}$, with the usual extension to $q=\infty$. 
For $A\subset\{1,\ldots,p\}$, $\bv_A=(v_j,j\in A)^T$ and $\bX_A=(\bx_k, k\in A)$, 
including $A = -j =\{1,\ldots,p\}\setminus \{j\}$. 

\subsection{Bias corrected linear estimators} 
In the classical theory of linear models, the least squares estimator of 
an estimable regression coefficient $\beta_j$ can be written as 
\bel{LSE}
\hbeta_j^{(lse)} := (\bx_j^\perp)^T \by /(\bx_j^\perp)^T\bx_j, 
\eel
where $\bx_j^\perp$ is the projection of $\bx_j$ to the orthogonal complement of the 
column space of $\bX_{-j} = (\bx_k, k\neq j)$. 
Since this is equivalent to solving the equations
$(\bx_j^\perp)^T(\by - \beta_j \bx_j)=(\bx_j^\perp)^T\bx_k=0$ $\forall\ k\neq j$ 
in the score system $\bv\to (\bx_j^\perp)^T\bv$, 
$\bx_j^\perp$ can be viewed as the score vector for the least squares estimation of $\beta_j$. 
For estimable $\beta_j$ and $\beta_k$, 
\bel{LSE-cov}
\Cov(\hbeta_j^{(lse)},\hbeta_k^{(lse)}) 
= \sigma^2(\bx_j^\perp)^T\bx_k^\perp/(\|\bx_j^\perp\|_2^2\,\|\bx_k^\perp\|_2^2). 
\eel

In the high-dimensional case $p>n$, $\rank(\bX_{-j})=n$ for all $j$ when $\bX$ is in general position.  
Consequently, $\bx_j^\perp=0$ and (\ref{LSE}) is undefined.  
However, it may still be interesting to preserve certain properties of the least squares estimator. 
This can be done by retaining the main equation $\bz_j^T(\by - \beta_j\bx_j)=0$ in a score system 
$\bz_j: \bv \to \bz_j^T\bv$ and relaxing the constraint $\bz_j^T\bx_k=0$ for $k\neq j$, resulting in a 
linear estimator. 
One advantage of (\ref{LSE}) is the explicit formula (\ref{LSE-cov}) for the covariance structure. 
This feature holds for all linear estimators of $\bbeta$. 
For any score vector $\bz_j$ not orthogonal to $\bx_j$, the corresponding 
univariate linear regression estimator satisfies
\bes
\hbeta_j^{(lin)} = \frac{\bz_j^T\by}{\bz_j^T\bx_j} = \beta_j + \frac{\bz_j^T\bep}{\bz_j^T\bx_j} 
+ \sum_{k\neq j}\frac{\bz_j^T\bx_k\beta_k}{\bz_j^T\bx_j}
\ees
with a similar covariance structure to (\ref{LSE-cov}). 
A problem with this linear estimator is its bias. 
For every $k\neq j$ with $\bz_j^T\bx_k\neq 0$, the contribution of $\beta_k$ to the bias is linear in $\beta_k$. 
Thus, under the assumption of $\|\bbeta\|_0\le 2$, which is very strong, 
the bias of $\hbeta_j^{(lin)}$ is still unbounded when $\bz_j^T\bx_k\neq 0$ for at least one $k\neq j$.  
We note that for $\rank(\bX_{-j})=n$, it is impossible to have $\bz_j\neq 0$ and 
$\bz_j^T\bx_k=0$ for all $k\neq j$, so that bias is unavoidable. 
Still, this analysis of the linear estimator suggests a bias correction  
with a nonlinear initial estimator $\hbbeta^{(init)}$: 
\bel{LDPE}
\hbeta_j = \hbeta_j^{(lin)}  
- \sum_{k\neq j}\frac{\bz_j^T\bx_k\hbeta^{(init)}_k}{\bz_j^T\bx_j}
=\frac{\bz_j^T\by}{\bz_j^T\bx_j} 
- \sum_{k\neq j}\frac{\bz_j^T\bx_k\hbeta^{(init)}_k}{\bz_j^T\bx_j}. 
\eel
One may also interpret (\ref{LDPE}) as a one-step self bias correction from the initial estimator and write
\bes
\hbeta_j := \hbeta_j^{(init)} + \frac{\bz_j^T\{\by-\bX\hbbeta^{(init)}\}}{\bz_j^T\bx_j}.
\ees
The estimation error of (\ref{LDPE}) can be decomposed as a sum of the noise and the approximation errors: 
\bel{error-decomp}
\hbeta_j - \beta_j = \frac{\bz_j^T\bep}{\bz_j^T\bx_j} 
+ \frac{1}{\bz_j^T\bx_j}\sum_{k\neq j}\bz_j^T\bx_k(\beta_k-\hbeta_k^{(init)}). 
\eel
We require that $\bz_j$ be a vector depending on $\bX$ only, 
so that $\bz_j^T\bep/\|\bz_j\|_2\sim N(0,\sigma^2)$. 
A full description of (\ref{LDPE}) still requires the  specification of the score vector $\bz_j$ 
and the initial estimator $\hbbeta^{(init)}$. These choices will be discussed in 
the following two subsections. 

\subsection{Low-dimensional projections} 
We propose to use as $\bz_j$ a relaxed orthogonalization of $\bx_j$ against other design vectors. 
Recall that $\bz_j$ aims to play the role of $\bx_j^\perp$, 
the projection of $\bx_j$ to the orthogonal complement of the column space of $\bX_{-j}=(\bx_k, k\neq j)$. 
In the trivial case where $\|\bx_j^\perp\|_2$ is not too small, we may simply take $\bz_j=\bx_j^\perp$. 
In addition to the case of $\rank(\bX_{-j})=n$, where $\bx_j^\perp=0$, a relaxed projection  
could be useful when $\|\bx_j^\perp\|_2$ is positive but small.  
Since a relaxed projection $\bz_j$ is used and the estimator (\ref{LDPE}) is 
a bias-corrected projection of $\by$ to the direction of $\bz_j$, 
hereafter we call (\ref{LDPE}) the low-dimensional projection estimator (LDPE) 
for easy reference. 

A proper relaxed projection $\bz_j$ should control both the noise and approximation error terms in 
(\ref{error-decomp}), given suitable conditions on $\{\bX,\bbeta\}$ and an initial estimator $\hbbeta^{(init)}$. 
By (\ref{error-decomp}), the approximation error of (\ref{LDPE}) can be bounded by 
\bel{bias-sum}
\Big|\sum_{k\neq j}\bz_j^T\bx_k(\beta_k-\hbeta_k^{(init)})\Big| 
\le \Big(\max_{k\neq j}\big|\bz_j^T\bx_k\big|\Big)\|\hbbeta^{(init)}-\bbeta\|_1. 
\eel
This conservative bound is conveniently expressed as the product of a known function of $\bz_j$ 
and the initial estimation error independent of $j$. 
For score vectors $\bz_j$, define 
\bel{eta-tau}
\eta_j = \max_{k\neq j}\big|\bz_j^T\bx_k\big|/\|\bz_j\|_2,\quad \tau_j = \|\bz_j\|_2/|\bz_j^T\bx_j|. 
\eel
We refer to $\eta_j$ as the bias factor since $\eta_j\|\hbbeta^{(init)}-\bbeta\|_1$ controls 
the approximation error in (\ref{bias-sum}) relative to the length of the score vector. 
We refer to $\tau_j$ as the noise factor, since $\tau_j\sigma$ is the standard deviation of the 
noise component in (\ref{error-decomp}). 
Since $\bz_j^T\bep\sim N(0,\sigma^2 \|\bz_j\|_2^2)$, (\ref{error-decomp}) yields
\bel{xi-tau}
\eta_j\|\hbbeta^{(init)}-\bbeta\|_1/\sigma = o(1)
\ \Rightarrow\ 
\tau_j^{-1}\big(\hbeta_j - \beta_j\big) \approx N(0,\sigma^2). 
\eel
Thus, we would like to pick a $\bz_j$ with a small $\eta_j$ for the asymptotic normality and 
a small $\tau_j$ for estimation efficiency. 
Confidence intervals for $\beta_j$ and linear functionals of them can be constructed provided 
the condition in (\ref{xi-tau}) and a consistent estimator of $\sigma$. 

We still need a suitable $\bz_j$, a relaxed orthogonalization of $\bx_j$ against other design vectors. 
When the unrelaxed $\bx_j^\perp$ is nonzero, 
it can be viewed as the residual of the least squares fit of $\bx_j$ on $\bX_{-j}$. 
A familiar relaxation of the least squares method is to add an $\ell_1$ penalty. 
This leads to the choice of $\bz_j$ as the residual of the Lasso.
Let $\hbgamma_j$ be the vector of coefficients from the Lasso regression of $\bx_j$ on $\bX_{-j}$. 
The Lasso-generated score is 
\bel{Lasso-z_j}
\bz_j = \bx_j - \bX_{-j}\hbgamma_j,\ 
\hbgamma_j = \argmin_{\bb}\Big\{\frac{\|\bx_j-\bX_{-j}\bb\|_2^2}{2n}+\lam_j \|\bb\|_1\Big\}. 
\eel
It follows from the Karush-Kuhn-Tucker conditions for (\ref{Lasso-z_j}) that 
$|\bx_k^T\bz_j/n|\le\lam_j$ for all $k \neq j$, 
so that (\ref{eta-tau}) holds with $\eta_j\le n\lam_j/\|\bz_j\|_2$. 
This gives many choices of $\bz_j$ with different $\{\eta_j,\tau_j\}$. 
Explicit choices of such a $\bz_j$, or equivalently a $\lam_j$, are described in the next subsection. 
A rationale for the use of a common penalty level $\lam_j$ for all components of $\bb$ 
in (\ref{Lasso-z_j}) is the standardization of all design vectors. 
In an alternative in Subsection 2.3 called the restricted LDPE (R-LDPE), the penalty 
is set to zero for certain components of $\bb$ in (\ref{Lasso-z_j}). 


\subsection{Specific implementations} 
We have to pick $\hbbeta^{(init)}$, $\hsigma$, and the $\lam_j$ in (\ref{Lasso-z_j}). 
Since consistent estimation of $\sigma$ and fully automatic choices of $\lam_j$ are needed, 
we use methods based on the scaled Lasso and the least squares estimator 
in the model selected by the scaled Lasso ({scaled Lasso-LSE}). 

The scaled Lasso \cite{Antoniadis10,SunZ10,SunZ11} is a joint convex minimization method given by
\bel{scaled-Lasso}
\big\{\hbbeta^{(init)},\hsigma\big\} = \argmin_{\bb,\sigma}
\Big\{\frac{\|\by-\bX\bb\|_2^2}{2\sigma n} + \frac{\sigma}{2}+\lam_0\|\bb\|_1\Big\},
\eel
with a preassigned penalty level $\lam_0$. 
This automatically provides an estimate of the noise level in addition to the initial estimator of $\bbeta$. 
We use $\lam_0=\lam_{univ}=\sqrt{(2/n)\log p}$ 
in our simulation study. 
Existing error bounds for the estimation of both $\bbeta$ and $\sigma$ require 
$\lam_0=A\sqrt{(2/n)\log(p/\eps)}$ with certain $A>1$ and $0<\eps\le 1$ \cite{SunZ11}. 

The estimator (\ref{scaled-Lasso}) has appeared in the literature in different forms. 
The joint minimization formulation was given in \cite{Antoniadis10}, 
and an equivalent algorithm in \cite{SunZ10}.
If the minimum over $\bb$ is taken first in (\ref{scaled-Lasso}), 
the resulting $\hsigma$ appeared earlier in \cite{Zhang10-mc+}. 
The square root Lasso \cite{BelloniCW11} gives the same $\bbeta^{(init)}$ with a different formulation, 
but not joint estimation. The formulations in \cite{Zhang10-mc+} and \cite{SunZ10} allow concave 
penalties and a degrees of freedom adjustment. 

The Lasso is biased, as is the scaled Lasso. 
Let $\Shat^{(init)}$ be the set of nonzero estimated coefficients by the scaled Lasso. 
When $\Shat^{(init)}$ catches most large {$|\beta_j|$}, the bias of (\ref{scaled-Lasso}) can be reduced 
by the least squares estimator in the selected model $\Shat^{(init)}$: 
\bel{lse-after}
\big\{\hbbeta^{(init)},\hsigma\big\} = \argmin_{\bb,\sigma}
\Big\{\frac{\|\by-\bX\bb\|_2^2}{2\sigma (n-|\Shat^{(init)}|)} + \frac{\sigma}{2}: 
b_j=0\ \forall\ j\not\in\Shat^{(init)} \Big\}. 
\eel
This defines the {scaled Lasso-LSE}. 
We use the same notation in (\ref{scaled-Lasso}) and (\ref{lse-after}) since they both give 
initial estimates for the LDPE (\ref{LDPE}) and a noise level estimator for statistical inference based on 
the LDPE. The specific estimators will henceforth be referred to by their names or 
as (\ref{scaled-Lasso}) and (\ref{lse-after}).  
The {scaled Lasso-LSE} enjoys similar analytical error bounds as the scaled Lasso and outperformed 
scaled Lasso in a simulation study \cite{SunZ11}. 

The scaled Lasso can be also used to determine $\lam_j$ for the $\bz_j$ in (\ref{Lasso-z_j}). 
However, the penalty level for the scaled Lasso, set to guarantee performance bounds for the 
estimation of regression coefficients and noise level, may not be the best for 
controlling the bias and the standard error of the LDPE. 
By  (\ref{eta-tau}) and (\ref{xi-tau}), it suffices to find a $\bz_j$ with small bias factor 
$\eta_j$ and small noise factor $\tau_j$. These quantities are always available. 
This is quite different from the estimation of $\{\bbeta,\sigma\}$ in (\ref{scaled-Lasso}) 
where the effect of over-fitting is unobservable. 

We choose $\lam_j$ by tracking $\eta_j$ and $\tau_j$ in the Lasso path. 
One of our ideas is to reduce $\eta_j$ by allowing some over fitting of $\bx_j$ 
as long as $\tau_j$ is reasonably small. 
Ideally, this slightly more conservative approach will lead to confidence intervals with more 
accurate coverage probability. Along the Lasso path for regressing $\bx_j$ against $\bX_{-j}$, let 
\bel{Lasso-path-j}
&& \hbgamma_j(\lam) = \argmin_{\bb}
\Big\{\|\bx_j-\bX_{-j}\bb\|_2^2/(2n) + \lam\|\bb\|_1\Big\},\ 
\\ \nonumber && \bz_j(\lam) = \bx_j-\bX_{-j}\hbgamma_j(\lam),\quad 
\\ \nonumber && \eta_j(\lam) = \max_{k\neq j}|\bx_k^T\bz_j(\lam)|/\|\bz_j(\lam)\|_2,  
\\ \nonumber &&\tau_j(\lam) = \|\bz_j(\lam)\|_2/|\bx_j^T\bz_j(\lam)|, 
\eel
be the coefficient estimator $\hbgamma_j$, residual $\bz_j$, the bias factor $\eta_j$, and 
the noise factor $\tau_j$, as functions of $\lam$.  
We compute $\bz_j$ according to the algorithm in Table \ref{table:alg}. 

\begin{table}
\caption{Computation of $\bz_j$ from the Lasso (\ref{Lasso-path-j})}
\begin{tabular}{cl}
\toprule
Input: & an upper bound $\eta_j^*$ for the bias factor, with default value $\eta^*_j=\sqrt{2\log p}$, \\
& tuning parameters $\kappa_0\in [0,1]$ and $\kappa_1\in (0,1]$; \\
Step 1: & (verify/adjust $\eta_j^*$ and compute the corresponding noise factor $\tau_j^*$) \\
& If $\eta_j(\lam)>\eta_j^*$ for all $\lam>0$, $\eta_j^* \leftarrow (1+\kappa_1)\inf_{\lam>0}\eta_j(\lam)$; \\
& $\lam \leftarrow \max\{\lam: \eta_j(\lam)\le\eta^*_j\}$, 
$\eta_j^*\leftarrow \eta_j(\lam)$, $\tau_j^*\leftarrow \tau_j(\lam)$; \\
Step 2: & (further reduction of the bias factor)\\
& $\lam_j \leftarrow \min\{\lam: \tau_j(\lam)\le (1+\kappa_0)\tau^*_j\}$; 
\\ 
Output: & $\lam_j$, $\bz_j\leftarrow \bz_j(\lam_j)$, $\tau_j\leftarrow \tau_j(\lam_j)$, $\eta_j\leftarrow \eta_j(\lam_j)$\\
\bottomrule
\addlinespace
\end{tabular}
\label{table:alg}
\end{table}

In Table \ref{table:alg}, Step 1 finds a feasible upper bound $\eta_j^*$ for 
the bias factor and the corresponding noise factor $\tau^*_j$.  
Step 2 seeks $\bz_j=\bz_j(\lam_j)$ in (\ref{Lasso-path-j}) at a certain level $\lam=\lam_j$ 
with a smaller $\eta_j=\eta_j(\lam_j)$, subject to the constraint 
$\tau(\lam_j)\le (1+\kappa_0)\tau_j^*$ on the noise factor.  
It follows from Proposition \ref{prop-1} (i) below that $\eta_j(\lam)$ is non-decreasing in $\lam$, 
so that searching for the smallest $\eta_j(\lam)$ is equivalent to searching for the smallest $\lam$ 
in Step 2, subject to the constraint. 

In the search for $\bz_j$ with smaller $\eta_j$ in Step 2, 
the relative increment in the noise factor $\tau_j$ is no greater than $\kappa_0$. 
This corresponds to a loss of relative efficiency no greater than $1-1/(1+\kappa_0)^2$ 
for the estimation of $\beta_j$. 
In our simulation experiments, $\kappa_0=1/4$ provides a suitable choice, 
compared with $\kappa_0=0$ and $\kappa_0=1/2$. 
We would like to emphasize here that the score vectors $\bz_j$ 
computed by the algorithm in Table \ref{table:alg} are completely determined by the design $\bX$.  

A main objective of the algorithm in Table \ref{table:alg} is to find a $\bz_j$ with a bias factor 
$\eta_j\le C\sqrt{\log p}$ 
to allow a uniform bias bound via (\ref{bias-sum}), (\ref{eta-tau}), and (\ref{xi-tau}).
It is ideal if $C=\sqrt{2}$ is attainable, but a reasonably small $C$ also works with the argument. 
When $\eta^*_j=\sqrt{2\log p}$ is not feasible, Step 1 finds a larger upper bound $\eta_j^*$ for the bias factor. 
When $\sup_\lam\eta_j(\lam)<\sqrt{2\log p}$, $\eta_j^*<\sqrt{2\log p}$ after the adjustment in Step 1, 
resulting in an even smaller $\eta_j$ in Step 2. 
This does happen in our simulation experiments. 
The choice of the target upper bound $\sqrt{2\log p}$ for $\eta_j$ is based on its feasibility 
as well as the sufficiency of $\eta_j\le \sqrt{2\log p}$ 
for the verification of the condition in (\ref{xi-tau}) 
based on the existing $\ell_1$ error bounds for the estimation of $\bbeta$. 
Proposition \ref{prop-1} below asserts that $\max_{j\le p}\eta^*_j\le C\sqrt{\log p}$ is feasible 
when $\bX$ allows an optimal rate of sparse recovery. 
In our simulation experiments, we are able to use $\eta^*_j\le\sqrt{2\log p}$ 
in all replications and settings for all variables, a total of more than 1 million instances. 
Moreover, the theoretical results in Subsection \ref{subsec-random-design} prove that 
for the $\eta_j^*$ in Table \ref{table:alg}, $\max_{j\le p}\eta^*_j\le 3\sqrt{\log p}$ 
with high probability under proper conditions on random $\bX$. 
It is worthwhile to note that both $\eta_j$ and $\tau_j$ are computed, and 
control of $\max_k\eta_k$ is not required for the LDPE to apply to variables with small $\eta_j$.

We have also experimented with an LDPE using a restricted Lasso relaxation for $\bz_j$. 
This R-LDPE (restricted LDPE) can be viewed as a special case of a more general 
weighted low dimensional projection with different levels of relaxation for 
different variables $\bx_k$ according to their correlation to $\bx_j$. 
Although we have used (\ref{bias-sum}) to bound the bias, 
the summands with larger absolute correlation $|\bx_j^T\bx_k/n|$ are likely to have a 
greater contribution to the bias due to {the} initial estimation error $|\hbeta^{(init)}_k-\beta_k|$. 
A remedy for this phenomenon is to force smaller $|\bz_j^T\bx_k/n|$ for large 
$|\bx_j^T\bx_k/n|$ with a weighted relaxation. 
For the Lasso (\ref{Lasso-z_j}), this weighted relaxation can be written as 
\bes
\bz_j = \bx_j - \bX_{-j}\hbgamma_j,\ 
\hbgamma_j = \argmin_{\bb}\Big\{\frac{\|\bx_j-\bX_{-j}\bb\|_2^2}{2n}
+\lam_j\sum_{k\neq j}w_k|b_k|\Big\},
\ees
with $w_k$ being a decreasing function of the absolute correlation $|\bx_j^T\bx_k/n|$. 
For the {R-LDPE}, we simply set $w_k=0$ for large $|\bx_j^T\bx_k/n|$ and $w_k=1$ for other $k$. 

Here is an implementation of the {R-LDPE}. 
Let $K_{j,m}$ be the index set of the $m$ largest $|\bx_j^T\bx_k|$ with $k\neq j$ 
and $\bP_{j,m}$ be the orthogonal projection to the linear span of $\{\bx_k, k\in K_{j,m}\}$. 
Let $\bz_j = f(\bx_j,\bX_{-j})$ denotes the algorithm in Table \ref{table:alg} as a mapping 
$(\bx_j,\bX_{-j})\to \bz_j$.  We compute the {R-LDPE} by taking the projection of all 
design vectors to the orthogonal complement of $\{\bx_k, k\in K_{j,m}\}$ before the 
application of the procedure in (\ref{Lasso-path-j}) and Table \ref{table:alg}. 
The resulting score vector can be written as 
\bel{z_j-2nd}
\bz_j = f(\bP_{j,m}^\perp\bx_j,\bP_{j,m}^\perp\bX_{-j}). 
\eel

We use the rest of this subsection to present some useful properties of the Lasso path (\ref{Lasso-path-j}) 
for the implementation of the algorithm in Table \ref{table:alg} and some sufficient conditions 
for the uniform bound $\max_j\eta^*_j\le C\sqrt{\log p}$ for the bias factors in the output. 
Let 
\bel{scaled-Lasso-j}
\hsigma_j(\lam) = \argmin_{\sigma}\min_{\bb}
\Big\{\frac{\|\bx_j-\bX_{-j}\bb\|_2^2}{2n\sigma} + \frac{\sigma}{2} + \lam\|\bb\|_1\Big\}
\eel
be the solution of $\hsigma$ in (\ref{scaled-Lasso}) 
with $\{\bX,\by,\lam_0\}$ replaced by $\{\bX_{-j},\bx_j,\lam\}$. 

\begin{proposition}\label{prop-1} 
(i) In the Lasso path (\ref{Lasso-path-j}), $\|\bz_j(\lam)\|_2$, $\eta_j(\lam)$, and $\hsigma_j(\lam)$ 
are nondecreasing functions of $\lam$, and $\tau_j(\lam)\le 1/\|\bz_j(\lam)\|_2$. Moreover, 
$\hbgamma_j(\lam)\neq 0$ implies $\eta_j(\lam)=\lam n/\|\bz_j(\lam)\|_2$. \\ 
(ii) Let $\lam_{univ}=\sqrt{(2/n)\log p}$. Then, 
\bel{prop-1-1}
\hsigma_j(C\lam_{univ})>0 \hbox{ iff } \{\lam>0: \eta_j(\lam) \le C\sqrt{2\log p}\}\neq \emptyset,
\eel 
and in this case, the algorithm in Table 1 provides 
\bel{prop-1-2}
& \eta_j\le \eta_j^*\le (1+\kappa_1 I_{\{C>1\}})(1\vee C)\sqrt{2\log p},\quad 
\tau_j\le n^{-1/2}(1+\kappa_0)/\hsigma_j(C\lam_{univ}).
\eel 
Moreover, when $\bz_j(0)=\bx_j^\perp =0$, 
$\eta_j(0+) \inf\{\|\bgamma_j\|_1: \bX_{-j}\bgamma_j=\bx_j\}=\sqrt{n}$.\\
(iii) Let $0<a_0<1\leq C_0<\infty$. Suppose that for $s=a_0n/\log p$
\bes
&& \inf_{\bdelta}\sup_{\bbeta}\Big\{\|\bdelta(\bX,\by) - \bbeta\|_2^2: \by=\bX\bbeta, 
\hbox{$\sum_{j=1}^p$} \min(|\beta_j|/\lam_{univ},1)\le s+1 \Big\}\le 2C_0s(\log p)/n.
\ees
Then, $\max_{j\le p}\eta_j^*\le (1+\kappa_1)\sqrt{(4C_0/a_0)\log p}$ for the algorithm in Table \ref{table:alg}. 
\end{proposition}

The monotonicity of $\|\bz_j(\lam)\|_2$ and $\eta_j(\lam)$ in Proposition \ref{prop-1} (i) 
provides directions of search in both steps of the algorithm in Table \ref{table:alg}. 

Proposition 1 (ii) provides mild conditions for controlling the bias factor at 
$\eta_j \le\eta^*_j\le C\sqrt{2\log p}$ and the standard error to the order $\tau_j=O(n^{-1/2})$. 
It asserts that $\eta_j^* \leq \sqrt{2\log p}$ 
when the scaled Lasso (\ref{scaled-Lasso-j}) with $\lambda = \lambda_{univ}$ yields a positive $\hsigma_j$. 
In the completely collinear case where $\bx_k=\bx_j$ for some $k\neq j$, 
$\inf\{\|\bgamma_j\|_1: \bx_j=\bX_{-j}\bgamma_j \} = 1$ gives the largest $\eta_j=\sqrt{n}$. 
This suggests a connection between the minimum feasible $\eta_j$ 
and certain ``near estimability'' of $\beta_j$, with small $\eta_j$ for nearly estimable $\beta_j$. 
It also provides a connection between the smallest $\eta_j(\lam)$ and an $\ell_1$ recovery problem, 
leading to Proposition \ref{prop-1} (iii). 

Proposition \ref{prop-1} (iii) asserts that the validity of the upper bound $\max_j\eta^*_j\le C\sqrt{\log p}$ 
for the bias factor is a consequence of the existence of an estimator $\bdelta$ with the $\ell_2$ 
recovery bound in the noiseless case of $\bep=0$. 
In the more difficult case of $\bep\sim N(0,\sigma^2\bI)$, $\ell_2$ error bounds of the same type 
have been proven under sparse eigenvalue conditions on $\bX$, 
and by Proposition \ref{prop-1} (iii), $\max_j\eta^*_j\le C\sqrt{\log p}$ is also a consequence of such conditions.

\subsection{Confidence intervals} 
In Section 3, we will provide sufficient conditions on $\bX$ and $\bbeta$ 
under which the approximation error 
in (\ref{error-decomp}) is of smaller order than the standard deviation of the noise 
component. 
We construct approximate confidence intervals for such configurations 
of $\{\bX,\bbeta\}$ as follows. 

The covariance of the noise component in (\ref{error-decomp}) is proportional to  
\bel{cov-LDPE}
\bV=(V_{jk})_{p\times p},\ \hbox{ where }\ 
V_{jk} = \frac{\bz_j^T\bz_k}{|\bz_j^T\bx_j||\bz_k^T\bx_k|}
= \sigma^{-2}\Cov\Big(\frac{\bz_j^T\bep}{\bz_j^T\bx_j}, \frac{\bz_k^T\bep}{\bz_k^T\bx_k}\Big). 
\eel
Let $\hbbeta = (\hbeta_1,\ldots,\hbeta_p)^T$ be the vector of LDPEs $\hbeta_j$ in (\ref{LDPE}). 
For sparse vectors $\ba$ with bounded $\|\ba\|_0$, 
e.g. $\|\ba\|_0=2$ for a contrast between two regression coefficients, 
an approximate $(1-\alpha)100\%$ confidence interval is 
\bel{LDPE-CI}
\big|\ba^T\hbbeta - \ba^T\bbeta\big| \le \hsigma\Phi^{-1}(1-\alpha/2)(\ba^T\bV\ba)^{1/2}, 
\eel
where $\Phi$ is the standard normal distribution function.
We may choose $\{\hbbeta^{(init)},\hsigma\}$ in (\ref{scaled-Lasso}) or (\ref{lse-after}) 
and $\bz_j$ in Table \ref{table:alg} or (\ref{z_j-2nd}) 
in the construction of $\hbbeta$ and the confidence intervals. 
An alternative, larger estimate of $\sigma$, producing more conservative approximate 
confidence intervals, is the penalized maximum likelihood estimator of \cite{StadlerBG10}. 

\section{Theoretical Results} 
{In this section, we prove that when the $\ell_1$ loss of the initial estimator $\hbbeta^{(init)}$ is of 
an expected magnitude and the noise level estimator $\hsigma$ is consistent, 
the LDPE based confidence interval has approximately the preassigned coverage probability 
for statistical inference of linear combinations of $\beta_j$ with sufficiently small $\eta_j$. Under proper conditions on 
$X$ such as those given in Proposition~\ref{prop-1}, the width of such confidence intervals 
is of the order $\tau_j\asymp n^{-1/2}$. 
The accuracy of the approximation for the coverage probability is sufficiently sharp to allow 
simultaneous interval estimation of all $\beta_j$ and sharp error bounds for the estimation 
and selection errors of thresholded LDPE. 
We use existing error bounds to verify the conditions on $\hbbeta^{(init)}$ and $\hsigma$ 
under a capped-$\ell_1$ relaxation of the sparsity condition $\|\bbeta\|_0\le s$, provided that $s\log p\ll n^{1/2}$. 
Random matrix theory is used in Subsection 3.4 to check regularity conditions. }

\subsection{Confidence {intervals} for preconceived parameters, deterministic design}
Here we establish the asymptotic normality of the LDPE (\ref{LDPE}) and the validity of 
the resulting confidence interval (\ref{LDPE-CI}) for a preconceived parameter. 
This result is new and useful in and of itself since high-dimensional data often present a few 
effects known to be of high interest in advance. 
Examples include treatment effects in clinical trials, or the effect of education 
on income in social-economical studies. 
Simultaneous confidence intervals for all individual $\beta_j$ and thresholded LDPE 
for the entire vector $\bbeta$ will be considered in the next subsection 
as consequences of this result. 

Let $\lam_{univ}=\sqrt{(2/n)\log p}$. 
Suppose (\ref{LM}) holds with a vector $\bbeta$ 
satisfying the following capped-$\ell_1$ sparsity condition:  
\bel{sparse-beta}
\hbox{$\sum_{j=1}^p$} \min\big\{|\beta_j|/(\sigma\lam_{univ}),1\big\} \le s. 
\eel
This condition holds if $\bbeta$ is $\ell_0$ sparse with $\|\bbeta\|_0\le s$ 
or $\ell_q$ sparse with $\|\bbeta\|_q^q/(\sigma\lam_{univ})^q\le s$, $0<q\le 1$. 
Let $\sigma^*=\|\bep\|_2/\sqrt{n}$.  
A generic condition we impose on the initial estimator is 
\bel{ell_1-err-bd}
P\Big\{ \|\hbbeta^{(init)} - \bbeta\|_1 \ge C_1 s \sigma^*\sqrt{(2/n)\log(p/\eps)}\Big\} \le \eps
\eel
for a certain fixed constant $C_1$ and all $\alpha_0/p^2 \le \eps \le 1$, 
where $\alpha_0\in (0,1)$ is a preassigned constant. 
We also impose a similar generic condition on an estimator $\hsigma$ for the noise level: 
\bel{sigma-err-bd}
P\Big\{ |\hsigma/\sigma^* - 1 | \ge C_2 s (2/n)\log(p/\eps) \Big\} \le \eps, \ 
\forall \alpha_0/p^2\le \eps\le 1, 
\eel 
with a fixed $C_2$. 
We use the same $\eps$ in (\ref{ell_1-err-bd}) and (\ref{sigma-err-bd}) without much loss of generality. 

By requiring fixed $\{C_1,C_2\}$, we implicitly impose regularity conditions 
on the design $\bX$ and the sparsity index $s$ in (\ref{sparse-beta}). 
Existing oracle inequalities can be used to verify (\ref{ell_1-err-bd}) for various regularized 
estimators of $\bbeta$ under different sets of conditions on $\bX$ and $\bbeta$ 
\cite{CandesT07,ZhangH08,BickelRT09,vandeGeerB09,
Zhang09-l1,Zhang10-mc+,YeZ10,SunZ11,ZhangZ11}.  
Although most existing results are derived for penalty/threshold levels depending on 
a known noise level $\sigma$ and under the $\ell_0$ sparsity condition on $\bbeta$, 
their proofs can be combined or extended to obtain (\ref{ell_1-err-bd}) once (\ref{sigma-err-bd}) 
becomes available. 
For the joint estimation of $\{\bbeta,\sigma\}$ with (\ref{scaled-Lasso}) or (\ref{lse-after}), 
specific sets of sufficient conditions for both (\ref{ell_1-err-bd}) and (\ref{sigma-err-bd}), 
based on \cite{SunZ11}, are stated in Subsection \ref{subsec-oracle}. 
In fact, the probability of the union of the two events is smaller than 
$\eps$ in {the} specific case where $\lam_0=A\sqrt{(2/n)\log(p/\eps)}$ 
in (\ref{scaled-Lasso}) for a certain $A>1$.

\begin{theorem}\label{th-1} 
Let $\hbeta_j$ be the LDPE in (\ref{LDPE}) with 
an initial estimator $\hbbeta^{(init)}$. 
Let $\eta_j$ and $\tau_j$ be the bias and noise factors in (\ref{eta-tau}), 
$\sigma^*=\|\bep\|_2/\sqrt{n}$, 
$\max(\eps_n',\eps_n'')\to 0$, and $\eta^*>0$. 
Suppose (\ref{ell_1-err-bd}) holds with $\eta^*C_1s\sqrt{(2/n)\log(p/\eps)}\le\eps_n'$. 
If $\eta_j\le\eta^*$, then
\bel{th-1-1}
P\Big\{ \big|\tau_j^{-1}(\hbeta_j - \beta_j) - \bz_j^T\bep/\|\bz_j\|_2\big| 
> \sigma^* \eps_n'\Big\}\le\eps. 
\eel
If in addition (\ref{sigma-err-bd}) holds with $C_2 s (2/n)\log(p/\eps)\le \eps_n''$, then for all 
$t\ge (1+\eps_n')/(1-\eps_n'')$, 
\bel{th-1-2}
P\Big\{|\hbeta_j - \beta_j| \ge \tau_j\hsigma t\Big\}\le 2\Phi_{n-1}(-(1-\eps_n'')t+\eps_n')+2\eps, 
\eel
where $\Phi_n(t)$ is the student-t distribution function with $n$ degrees of freedom. 
Moreover, for the covariance matrix $\bV$ in (\ref{cov-LDPE}) and all fixed $m$, 
\bel{th-1-3}\qquad
\lim_{n\to\infty} 
\inf_{\ba\in\scrA_{n,p,m}}P\Big\{\big|\ba^T\hbbeta - \ba^T\bbeta\big| 
\le \hsigma\Phi^{-1}(1-\alpha/2)(\ba^T\bV\ba)^{1/2}\Big\} = 1-\alpha, 
\eel
where $\Phi(t) = P\{ N(0,1)\le t\}$ and 
$\scrA_{n,p,m}=\{\ba
: \|\ba\|_0\le m, \max_{j\le p}|a_j|\eta_j\le\eta^*\}$.  
\end{theorem}

Since $(\bz_j^T\bep/\|\bz_j\|_2, j\le p)$ has a multivariate normal 
distribution with identical marginal distributions $N(0,\sigma^2)$, 
(\ref{th-1-1}) establishes the joint asymptotic normality of 
the LDPE for finitely many $\hbeta_j$ under (\ref{ell_1-err-bd}). 
This allows us to write the LDPE as an approximate Gaussian sequence 
\bel{Gaussian-seq}
\hbeta_j = \beta_j + N(0,\tau_j^2\sigma^2) + o_P(\tau_j\sigma). 
\eel
Under the additional condition (\ref{sigma-err-bd}), 
(\ref{th-1-2}) and (\ref{th-1-3}) justify  
the approximate coverage probability of the resulting confidence intervals. 

\begin{remark}\label{remark-unif-sig}
 In Theorem \ref{th-1}, all conditions on $\bX$ and $\bbeta$ are imposed through
(\ref{ell_1-err-bd}), (\ref{sigma-err-bd}), and the requirement of relatively small $\eta_j$ 
to work with these conditions. The uniform signal strength condition, 
\bel{unif-sig}
\hbox{$\min_{\beta_j\neq 0}$} |\beta_j| \ge C\sigma\sqrt{(2/n)\log p},\ C>1/2, 
\eel
required for variable selection consistency \cite{Wainwright09b, Zhang10-mc+}, 
is not required for (\ref{ell_1-err-bd}) and (\ref{sigma-err-bd}). 
This is the most important feature of the LDPE that sets {it} apart {from} variable selection approaches. 
More explicit sufficient conditions for (\ref{ell_1-err-bd}) and (\ref{sigma-err-bd}) 
are given in Subsection \ref{subsec-oracle} for the initial estimators (\ref{scaled-Lasso}) 
and (\ref{lse-after}). 
\end{remark}

\begin{remark}Although Theorem \ref{th-1} does not require $\tau_j$ to be small, 
the noise factor is proportional to the width of the confidence interval and thus 
its square is reciprocal to the efficiency of the LDPE. 
The bias factor $\eta_j$ is required to be relatively small for (\ref{th-1}) and (\ref{th-2}), 
but no condition is imposed on $\{\eta_k, k\neq j\}$ for the inference of $\beta_j$. 
Since $\eta_j$ and $\tau_j$ are computed in Table \ref{table:alg}, one may apply 
Theorem \ref{th-1} to a set of the easy-to-estimate $\beta_j$ with small $\{\eta_j,\tau_j\}$ 
and leave out some hard-to-estimate regression coefficients. 
\end{remark}

In our implementation in Table \ref{table:alg}, $\bz_j$ is the residual of the Lasso estimator 
in the regression model for $\bx_j$ against $\bX_{-j}=(\bx_k,k\neq j)$. 
It follows from Proposition \ref{prop-1} that under proper conditions on the design matrix, 
$\eta_j\asymp \sqrt{\log p}$ and $\tau_j\le 1/\|\bz_j\|_2 \asymp n^{-1/2}$ for the algorithm in 
Table \ref{table:alg}. Such rates are realized in the simulation experiments described in Section 4 
and further verified for Gaussian designs in Subsection \ref{subsec-random-design}.
Thus, the dimension constraint for the asymptotic normality and proper coverage 
probability in Theorem \ref{th-1} is $s(\log p)/\sqrt{n}\to 0$. 

\subsection{Simultaneous confidence intervals and the thresholded LDPE} 
Here we provide theoretical justifications for simultaneous applications of the proposed LDPE 
confidence interval, with multiplicity adjustments, in the absence of a preconceived parameter of interest. 
In Theorem~\ref{th-1}, (\ref{th-1-1}) is uniform in $\eps\in [\alpha_0/p^2,1]$ and 
(\ref{th-1-2}) is uniform in the corresponding $t$. This uniformity 
allows Bonferroni adjustments to control familywise error rate in 
simultaneous interval estimation. This uniformity also applies to the approximation in (\ref{Gaussian-seq}), 
leading to sharp $\ell_2$ and selection error bounds of a thresholded LDPE for the estimation 
of the entire vector $\bbeta$. 
We present these consequences of Theorem~\ref{th-1} in the following two theorems.

\begin{theorem}\label{cor-1} Suppose (\ref{ell_1-err-bd}) holds with 
$\eta^*C_1s \sqrt{(2/n)\log(p/\eps)}\le\eps_n'$. Then, 
\bel{cor-1-1}
P\Big\{ \max_{\eta_j\le \eta^*} \big|\tau_j^{-1}(\hbeta_j - \beta_j) - \bz_j^T\bep/\|\bz_j\|_2\big| 
> \sigma^* \eps_n'\Big\}\le\eps. 
\eel
If (\ref{sigma-err-bd}) also holds with $C_2 s (2/n)\log(p/\eps)\le \eps_n''$, then 
for all $j\le p$ and $t\ge (1+\eps_n')/(1-\eps_n'')$, 
\bel{cor-1-2}
P\Big\{ \max_{\eta_j\le \eta^*} |\hbeta_j - \beta_j|/(\tau_j\hsigma) > t\Big\}
\le 2\Phi_n( - (1-\eps_n'')t + \eps_n')\#\{j:\eta_j\le\eta^*\}+2\eps. 
\eel
If, in addition to (\ref{ell_1-err-bd}) and (\ref{sigma-err-bd}), 
$\max_{j\le p}\eta_j\le \eta^*$ and $\max(\eps_n',\eps)\to 0$ as $\min(n,p)\to \infty$, 
then 
for fixed $\alpha\in (0,1)$ and $c_0>0$, 
\bel{cor-1-3}
\liminf_{n\to \infty} P\Big\{ \max_{j\le p} \Big|\frac{\hbeta_j - \beta_j}{\tau_j(\hsigma\wedge \sigma)}\Big|  
\le  c_0 + \sqrt{2\log(p/\alpha)}\Big\}  \ge 1-\alpha. 
\eel
\end{theorem}

The error bound (\ref{cor-1-1}) asserts that the $o_P(1)$ in (\ref{Gaussian-seq}) is uniform 
in $j$. This uniform central limit theorem and the simultaneous confidence intervals 
(\ref{cor-1-2}) and (\ref{cor-1-3}) are valid as long as (\ref{ell_1-err-bd}) and (\ref{sigma-err-bd}) hold 
with $s\log p = o(n^{1/2})$. Since (\ref{ell_1-err-bd}) and (\ref{sigma-err-bd}) are consequences 
of (\ref{sparse-beta}) and proper regularity conditions on $\bX$, these results do not 
require the uniform signal strength condition (\ref{unif-sig}).

It follows from (\ref{Gaussian-seq}) and Proposition \ref{prop-1} (ii) that for a fixed $j$, 
the estimation error of $\hbeta_j$ is of the order $\tau_j\sigma$ and 
$\tau_j\asymp n^{-1/2}$ under proper conditions.  With penalty level 
$\lam=\sigma\sqrt{(2/n)\log p}$, the Lasso may have a high probability of estimating 
$\beta_j$ by zero when $\beta_j = \lam/2$. Thus, in the worst case scenario, the Lasso 
inflates the error by a factor of order $\sqrt{\log p}$. Of course, the Lasso is super 
efficient when it estimates the actual zero $\beta_j$ by zero. 

The situation is different for the estimation of the entire vector $\bbeta$. 
The raw LDPE has an $\ell_2$ error of order $\sigma^2p/n$, compared with 
$\sigma^2 s(\log p)/n$ for the Lasso. However, this is not what the LDPE is designed for. 
The {thrust} of the LDPE approach is to turn the regression problem (\ref{LM}) into a 
Gaussian sequence model (\ref{Gaussian-seq}) with uniformly small approximation error 
and a consistent estimator of the covariance structure. The raw LDPE is sufficient for 
statistical inference of a preconceived $\beta_j$. For the estimation of the entire $\bbeta$ 
or variable selection, our recommendation is to use a thresholded LDPE. 
{We may use either the hard or soft thresholding methods: 
\bel{thresh}
\hbeta_j^{(thr)} &=& \begin{cases}
\hbeta_j I\{|\hbeta_j| > \that_j\}, & \hbox{(hard threshold)}\cr
\sgn(\hbeta_j)\big(|\hbeta_j| - \that_j\big)^+, & \hbox{(soft threshold),}\end{cases}
\\ \nonumber \Shat^{(thr)} &=& \{j: |\hbeta_j| > \that_j\},
\eel
where $\hbeta_j$ is as in Theorem~\ref{th-1} and
$\that_j \approx \hsigma\tau_j\Phi^{-1}(1-\alpha/(2p))$ with $\alpha>0$. 
Although the theory is similar between the two \cite{DonohoJ94}, our explicit analysis focuses on 
soft-thresholding. }

\begin{theorem}\label{cor-2} 
Let $L_0=\Phi^{-1}(1-\alpha/(2p))$, $\ttil_j = \tau_j\sigma L_0$, and 
$\that_j = (1+{c}_n)\hsigma\tau_jL_0$ with positive constants $\alpha$ and ${c}_n$. Suppose 
(\ref{ell_1-err-bd}) holds with $\eta^*C_1s/\sqrt{n}\le\eps_n'$, $\max_{j\le p}\eta_j\le\eta^*$, and 
\bel{cor-2-0}
P\Big\{\frac{(\hsigma/\sigma)\vee(\sigma/\hsigma) -1+\eps_n'\sigma^*/(\hsigma\wedge \sigma)}
{1-(\hsigma/\sigma-1)_+ } > {c}_n\Big\} \le 2\eps. 
\eel
Let $\hbbeta^{(thr)} = (\hbeta_1^{(thr)},\ldots,\hbeta_p^{(thr)})^T$ be the {soft} thresholded LDPE (\ref{thresh}) 
with these $\that_j$. 
Then, there exists an event $\Omega_n$ with $P\{\Omega_n^c\}\le 3\eps$ such that 
\bel{cor-2-1}
&&E\|\hbbeta^{( thr)} - \bbeta\|_2^2I_{\Omega_n}
\le \sum_{j=1}^p \min\Big\{\beta_j^2,\tau_j^2\sigma^2(L_0^2(1+2{c}_n)^2+1)\Big\}
+(\eps L_n/p)\sigma^2\sum_{j=1}^p\tau_j^2, 
\eel
where $L_n = 4/L_0^3+4{c}_n/L_0+12{c}_n^2L_0$. 
Moreover, with at least probability $1-\alpha - 3\eps$, 
\bel{cor-2-2}
\{j: |\beta_j|> (2+2{c}_n)\ttil_j\} \subseteq \Shat^{(thr)} \subseteq \{j:\beta_j\neq 0\}. 
\eel
\end{theorem}

Theorem \ref{cor-2} asserts that thresholding the LDPE provides similar error bounds to 
thresholding a Gaussian sequence $N(\beta_j,\tau_j^2\sigma^2), j\le p$. 
Since $\max_{j\le p}\eta_j\le C\sqrt{\log p}$ can be achieved 
under mild conditions, the main requirement is $s\sqrt{(\log p)/n}\to 0$ 
for the estimation and selection error bounds in (\ref{cor-2-1}) and (\ref{cor-2-2}). 
This is a weaker requirement than $s(\log p)/\sqrt{n}\to 0$ for the asymptotic normality in (\ref{th-1-3}). 
When $C_2s(2L_0^2/n)\le \eps_n''$, (\ref{cor-2-0}) follows from (\ref{sigma-err-bd}) and 
$P\{ (1-\eps_n''')/(1-\eps_n'') \le \sigma^*/\sigma \le (1+\eps_n''')/(1+\eps_n'')\}\le\eps$ with 
${c}_n\ge (\eps_n'''+\eps_n')/(1- \eps_n''')^2$. 
The condition on $\sigma^*/\sigma$ is easy to check since $(\sigma^*/\sigma)^2\sim \chi^2_n/n$. 
In what follows, we always assume that proper small constants ${c}_n>0$ are taken in 
(\ref{cor-2-0}) so that it is a consequence of (\ref{sigma-err-bd}).

\begin{remark}
The major difference between (\ref{cor-2-2}) and the existing variable selection consistency theory 
is again in the signal requirement.  
Variable selection consistency requires the uniform signal strength condition 
(\ref{unif-sig}) as discussed in Remark \ref{remark-unif-sig}, and 
existing variable selection methods are not guaranteed to select correctly variables 
with large $|\beta_j|$ or $\beta_j=0$ in the presence of small $|\beta_j|\neq 0$. 
In comparison, Theorem \ref{cor-2} makes no assumption of (\ref{unif-sig}). 
Under the regularity conditions for (\ref{cor-2-2}), large $|\beta_j|$ are selected by the thresholded 
LDPE and $\beta_j=0$ are not selected, in the presence of possibly many small nonzero $|\beta_j|$. 
\end{remark}

The analytical difference between the thresholded LDPE and existing regularized estimators 
lies in the quantities thresholded. 
For the LDPE, the effect of thresholding to the approximate Gaussian sequence (\ref{Gaussian-seq}) 
is explicit and requires only univariate analysis to understand. 
In comparison, for the Lasso and some other regularized estimators, thresholding is applied to 
the gradient $\bX^T(\by-\bX\hbbeta)/n$ via the Karush-Kuhn-Tucker type condition, 
leading to more complicated nonlinear multivariate analysis.

For the estimation of $\bbeta$, the order of the $\ell_2$ error bound in (\ref{cor-2-1}), 
$\sum_{j=1}^p \min(\beta_j^2, \sigma^2\lam_{univ}^2)$, 
is slightly sharper than the typical order of $\|\bbeta\|_0\sigma^2\lam_{univ}^2$ or 
$\sigma\lam_{univ}\sum_{j=1}^p \min\big\{|\beta_j|, \sigma\lam_{univ}\big\}$ in the literature, 
where $\lam_{univ}=\sqrt{(2/n)\log p}$. 
However, since the Lasso and other regularized estimators are proven to be rate optimal in the 
$\ell_2$ estimation loss for many classes {of} sparse $\bbeta$, 
the main advantage of the thresholded LDPE seems to be the clarity of the effect of thresholding 
to the individual $\hbeta_j$ in the approximate Gaussian sequence (\ref{Gaussian-seq}).

\subsection{Checking conditions by oracle inequalities.}\label{subsec-oracle}
Our main theoretical results, stated in Theorems \ref{th-1}, \ref{cor-1}, and \ref{cor-2} 
in the above two subsections, 
provide justifications for the LDPE-based confidence interval of a single preconceived linear parameter 
of $\bbeta$, simultaneous confidence intervals for all $\beta_j$, and the estimation and selection error 
bounds for the thresholded LDPE for the vector $\bbeta$. 
These results are based on conditions (\ref{ell_1-err-bd}) and (\ref{sigma-err-bd}). 
We have mentioned that for proper $\hbbeta^{(init)}$ and $\hsigma$, 
these two generic conditions can be verified in many ways under condition (\ref{sparse-beta}) 
based on existing results. 
The purpose of this subsection is to describe a specific way of verifying these two conditions 
and thus provide a more definitive and complete version of the theory. 

In regularized linear regression, oracle inequalities have been established 
for different regularized estimators and loss functions. 
We confine our discussion here to the scaled Lasso (\ref{scaled-Lasso}) and {the scaled Lasso-LSE} 
(\ref{lse-after}) as specific choices of the initial estimator, since the confidence 
interval in Theorem \ref{th-1} is based on the joint estimation of regression coefficients and the noise level. 
We further confine our discussion to bounds for the $\ell_1$ error of $\hbbeta^{(init)}$ and 
the relative error of $\hsigma$ 
involved in (\ref{ell_1-err-bd}) and (\ref{sigma-err-bd}). 

We use the results in \cite{SunZ11} where properties of estimators (\ref{scaled-Lasso}) and (\ref{lse-after}) 
were established based on a compatibility factor \cite{vandeGeerB09} and sparse eigenvalues.  
Let $\xi\ge 1$, $S=\{j: |\beta_j|>\sigma \lam_{univ}\}$, and 
$\scrC(\xi,S) = \{\bu: \|\bu_{S^c}\|_1\le \xi\|\bu_S\|_1\}$. 
The compatibility factor is defined as 
\bel{kappa}
\kappa(\xi,S) = \inf\big\{\|\bX\bu\|_2|S|^{1/2}/(n^{1/2}\|\bu_S\|_1): 0\neq \bu\in\scrC(\xi,S)\big\}. 
\eel
Let $\phi_{\min}$ and $\phi_{\max}$ denote the smallest and largest eigenvalues of matrices respectively. 
For positive integers $m$, define sparse eigenvalues as 
\bel{sparse-eigen}
&& \phi_-(m,S)=\min_{B\supset S,|B\setminus S|\le m}\phi_{\min}(\bX_B^T\bX_B/n),\ 
\cr && \phi_+(m,S)=\min_{B\cap S=\emptyset,|B|\le m}\phi_{\max}(\bX_B^T\bX_B/n). 
\eel
The following theorem is a consequence of checking the conditions of Theorem \ref{th-1} 
by Theorems 2 and 3 in \cite{SunZ11}. 

\begin{theorem}\label{th-2} 
Let $\{A,\xi,{c}_0\}$ be fixed positive constants with $\xi>1$ and $A > (\xi+1)/(\xi-1)$. 
Let $\lam_0=A\sqrt{(2/n)\log(p/\eps)}$. 
Suppose $\bbeta$ is sparse in the sense of (\ref{sparse-beta}), 
$\kappa^2(\xi,S)\ge {c}_0$, and $(s\vee 1)(2/n)\log(p/\eps)\le \mu_*$ for a certain $\mu^*>0$. \\
(i) Let $\hbbeta^{(init)}$ and $\hsigma$ be the scaled Lasso estimator in (\ref{scaled-Lasso}).  
Then, conditions (\ref{ell_1-err-bd}) and (\ref{sigma-err-bd}) hold for certain constants 
$\{\mu_*,C_1,C_2\}$ depending on $\{A,\xi,{c}_0\}$ only. 
Consequently, all conclusions of Theorems \ref{th-1}, \ref{cor-1}, and \ref{cor-2} hold with
$C_1\eta^*(s\lam_0/A)\le \eps_n'$ and $C_2s(\lam_0/A)^2 \le \eps_n''$. \\
(ii) Let $\hbbeta^{(init)}$ and $\hsigma$ be the {scaled Lasso-LSE} in (\ref{lse-after}). 
Suppose $\xi^2/\kappa^2(\xi,S) \le K/\phi_+(m,S)$ and $\phi_-(m,S)\ge {c}_1>0$ for certain $K>0$ and 
integer $m-1< K|S|\le m$. 
Then, (\ref{ell_1-err-bd}) and (\ref{sigma-err-bd}) hold for certain constants $\{\mu^*,C_1,C_2\}$ 
depending on $\{A,\xi,{c}_0,{c}_1,K\}$ only. 
Consequently, 
all conclusions of Theorems \ref{th-1}, \ref{cor-1}, and \ref{cor-2} hold with
$C_1\eta^*(s\lam_0/A)\le \eps_n'$ and $C_2s(\lam_0/A)^2 \le \eps_n''$. 
\end{theorem}

\begin{remark}\label{remark-th2} 
Let $A>(\xi+1)/(\xi-1)$ as in Theorem \ref{th-2} (i). Then, there exist constants $\{\tau_0,\nu_0\}\subset (0,1)$ 
satisfying the condition $(1-\tau_0^2)A = (\xi+1)/\{\xi-(1+\nu_0)/(1-\nu_0)\}$. 
For these $\{\tau_0,\nu_0\}$, $n\ge 3$, and $p\ge 7$, we may take 
\bel{remark-th-2-1}
\mu_* = \min\Big\{\frac{2{c}_0\tau_0^2}{A^2(\xi+1)}, \frac{\tau_0^2/(1/\nu_0-1)}{2A(\xi+1)},\log(4/e)\Big\},\ 
C_2=\frac{\tau_0^2}{\mu_*},\ 
C_1=\frac{C_2}{A(1-\tau_0^2)}. 
\eel
\end{remark}

\medskip

The main conditions of Theorem \ref{th-2} are 
\bel{cond-th-2}
\kappa^2(\xi,S)\ge {c}_0,\ 
\xi^2/\kappa^2(\xi,S)\le K/\phi_+(m,S),\ \phi_-(m,S)\ge {c}_1, 
\eel 
where $m$ is the smallest integer upper bound of $K|S|$. 
While Theorem \ref{th-2} (i) requires only the first inequality in (\ref{cond-th-2}), 
Theorem \ref{th-2} (ii) requires all three.  Let 
\bes
&& RE_2(\xi,S) = \inf\big\{\|\bX\bu\|_2/(n^{1/2}\|\bu\|_2):  
\bu\in\scrC(\xi,S), u_j\bx_j^T\bX\bu\le 0, j\notin S \big\}, 
\cr &&F_1(\xi,S) = \inf\big\{\|\bX^T\bX\bu\|_\infty|S|/(n\|\bu_S\|_1):  
\bu\in\scrC(\xi,S), u_j\bx_j^T\bX\bu\le 0, j\notin S \big\}, 
\ees
be respectively the restricted eigenvalue and sign restricted cone invertibility factor for the 
Gram matrix. It is worthwhile to note that 
\bel{factors}
F_1(\xi,S)\ge \kappa^2(\xi,S) \ge RE^2_2(\xi,S)
\eel
always holds and lower bounds of these quantities can be expressed in terms of sparse eigenvalues \cite{YeZ10}. 
By \cite{SunZ11}, one may replace $\kappa^2(\xi,S)$ throughout Theorem \ref{th-2} with $F_1(\xi,S)$. 
In view of (\ref{factors}), this will actually weaken the condition. 
However, since more explicit proofs are given in terms of $\kappa(\xi,S)$ in \cite{SunZ11}, 
the compatibility factor is used in Theorem \ref{th-2} to facilitate {a direct matching} of proofs 
between the two papers.
By \cite{ZhangH08,Zhang10-mc+,HuangZ12},  (\ref{cond-th-2}) can be replaced by 
the sparse Riesz condition,  
\bel{src}
s \le d^*/\{\phi_+(d^*,\emptyset)/\phi_-(d^*,\emptyset)+1/2\}. 
\eel
Proposition~\ref{prop-2} below provides a way of 
checking (\ref{cond-th-2}) for a given design in (\ref{LM}). 

\begin{proposition}\label{prop-2} 
Let $\{\xi,M_0,{c}_*,c^*\}$ be fixed positive constants, $\lam_1 = M_0\sqrt{(\log p)/n}$,  and 
\bes
\hbSigma = \Big((\bx_j^T\bx_k/n)I\{|\bx_j^T\bx_k/n|\ge\lam_1\}\Big)_{p\times p}
\ees
be the thresholded Gram matrix. 
Suppose $\phi_{\min}(\hbSigma)\ge {c}_*$ and $s\lam_1(1+\xi)^2 \le {c}_*/2$. 
Then, for all $|S|\le s$, $\kappa^2(\xi,S)\ge {c}_*/2$. 
Let $K = 2\xi^2(c^*/{c}_*+1/2)$. 
If in addition, $\phi_{\max}(\hbSigma)\le c^*$ and $s\lam_1(1+K)+\lam_1 \le {c}_*/2$, then 
$\phi_-(m,S)\ge {c}_*/2$ and (\ref{cond-th-2}) holds with ${c}_0={c}_*/2$. 
\end{proposition}

The main condition of Proposition \ref{prop-2} is a small $s\sqrt{(\log p)/n}$. 
This is not restrictive since Theorem \ref{th-1} requires the stronger condition of a small $s(\log p)/\sqrt{n}$. 
It follows from \cite{BickelL08a} that after hard thresholding at a level of order $\lam_1$, sample covariance 
matrices converge to a population covariance matrix in the spectrum norm under mild sparsity conditions 
on the population covariance matrix. 
Since convergence in the spectrum norm implies convergence of the minimum and maximum eigenvalues, 
$\phi_{\min}(\hbSigma)\ge {c}_*$ and  $\phi_{\max}(\hbSigma)\le c^*$ are reasonable conditions. 
This and other applications of random matrix theory are discussed in the next subsection. 

\subsection{Checking conditions by random matrix theory}\label{subsec-random-design} 
The most basic conditions for our main theoretical results in Subsections 3.1 and 3.2 
are (\ref{ell_1-err-bd}), (\ref{sigma-err-bd}), and the existence of $\bz_j$ with small $\eta_j$ and $\tau_j$. 
For deterministic design matrices, sufficient conditions for (\ref{ell_1-err-bd}) and 
(\ref{sigma-err-bd}) are given in Theorem~\ref{th-2} in the form of (\ref{cond-th-2}), 
and sufficient conditions for the existence of $\eta_j \le C\sqrt{\log p}$ and $\tau_j\asymp n^{-1/2}$ 
are given in Proposition \ref{prop-1}. These sufficient conditions are all analytical ones on the 
design matrix. In this subsection, we use random matrix theory to check these conditions 
with more explicit constant factors. 

The conditions of Theorems \ref{th-1} and \ref{th-2} hold in the following classes of design 
matrices: 
\bel{class}
\scrX_{s,n,p} &=& \scrX_{s,n,p}({c}_*,\delta,\xi,K)
\cr &=&\Big\{\bX:  \max_{j\le p}\eta_j\le 3\sqrt{\log p},\  
\max_{j\le p}\tau_j^2\sigma_j^2\le 2/n,\ 
\min_{|S|\le s}\kappa^2(\xi,S)\ge {c}_*(1-\delta)/4, 
\cr &&\qquad\qquad \max_{|S|\le s}\phi_+(m,S)\xi^2/\kappa^2(\xi,S)\le K,\ 
\min_{|S|\le s}\phi_-(m,S)\ge {c}_*(1-\delta) \Big\}, 
\eel
for certain positive $\{s,{c}_*,\delta,\xi,K\}$, 
where $\{\eta_j,\tau_j\}$ are computed from $\bX$ by the algorithm in Table \ref{table:alg} with 
$\kappa_0 \le 1/4$ and $3/(1+\kappa_1) > \sqrt{8}$, and $\kappa(\xi,S)$ and $\phi_\pm(m,S)$ 
are the compatibility factor and sparse eigenvalues of $\bX$ given in (\ref{kappa})
and (\ref{sparse-eigen}), with $m-1 < Ks\le m$. We note that $1/\sigma_j^2 \le 1/{c}_*$ by (\ref{Theta}), 
so that $\max_{j\le p}\tau_j^2 \le 2/(n{c}_*)$ in $\scrX_{s,n,p}({c}_*,\delta,\xi,K)$.  

Let $P_{\bSigma}$ be probability measures under which 
\bel{P-measure}
\tbX = (\tbx_1,\ldots,\tbx_p)\in\R^{n\times p}\ \hbox{ has iid $N(0,\bSigma)$ rows}. 
\eel
The column standardized version of $\tbX = (\tbx_1,\ldots,\tbx_p)$ is 
\bel{random-X}
\bX = (\bx_1,\ldots,\bx_p),\quad \bx_j=\tbx_j\sqrt{n}/\|\tbx_j\|_2. 
\eel
Since our discussion is confined to column standardized design matrices for simplicity, 
we assume without loss of generality that the diagonal elements of $\bSigma$ all equal to 1. 
Under $P_{\bSigma}$, $\bX$ does not have independent rows but $\bx_j$ is still related to $\bX_{-j}$ through 
\bel{LM-j}
\bx_j = \bX_{-j}\bgamma_j + \bep_j\sqrt{n}/\|\tbx_j\|_2,\ \bep_j\sim N(0,\sigma_j^2I_{n\times n}), 
\eel
where $\bep_j$ is independent of $\bX_{-j}$. 
Let $\Theta_{jk}$ be the elements of $\bSigma^{-1}$. 
Since the linear regression of $\tbx_j$ against $(\tbx_k,k\neq j)$ has coefficients $- \Theta_{jk}/\Theta_{jj}$ 
and noise level $1/\Theta_{jj}$, we have 
\bel{Theta}
\bgamma_j = \Big(-\sigma_j^2\Theta_{jk}\|\tbx_k\|_2/\|\tbx_j\|_2, k\neq j\Big)^T,\ 
\sigma_j^2=1/\Theta_{jj}. 
\eel

The aim of this subsection is to prove that $P_{\bSigma}(\scrX_{s,n,p})$ is uniformly large for a general 
collection of $P_{\bSigma}$. 
This result has two interpretations. 
The first interpretation is that when $\bX$ is indeed generated in accordance with 
(\ref{P-measure}) and (\ref{random-X}), the regularity conditions have a high probability to hold. 
The second interpretation is that $\scrX_{s,n,p}$, a deterministic subset of $\R^{n\times p}$, 
is sufficiently large as measured by $P_{\bSigma}$ in the collection. 
Since $\scrX_{s,n,p}$ does not depend on $\bSigma$ and the probability measures $P_{\bSigma}$  
are nearly orthogonal for different $\bSigma$, 
the use of $P_{\bSigma}$ does not add the random design assumption to our results. 

The following theorem specifies $\{{c}_*,c^*,\delta,\xi,K\}$ in (\ref{class}) for which 
$P_{\bSigma}\{\scrX_{s,n,p}({c}_*,\delta,\xi,K)\}$ is large when $s(\log p)/n$ is small. 
This works with the LDPE theory since $s(\log p)/\sqrt{n}\to 0$ is required anyway in Theorem \ref{th-1}. 
Define a class of coefficient vectors with small $\ell_q$ tail as 
\bes
\scrB_q(s,\lam) = \Big\{\bb\in\R^p: \hbox{$\sum_{j=1}^p$}\min(|b_j|^q/\lam^q,1) \le s \Big\}. 
\ees
We note that $\scrB_q(s,\sigma\lam_{univ})$ is the collection of all $\bbeta$ satisfying 
the capped-$\ell_1$ sparsity condition (\ref{sparse-beta}). 

\begin{theorem}\label{th-3} Suppose $\diag(\bSigma)=\bI_{p\times p}$, 
eigenvalues$(\bSigma)\subset [{c}_*,c^*]$, and all rows of $\bSigma^{-1}$ are in $\scrB_1(s,\lam_{univ})$. 
Then, there exist positive numerical constants $\{\delta_0,\delta_1,\delta_2\}$ and $K$ depending 
only on $\{\delta_1,\xi,{c}_*,c^*\}$ such that  
\bes
\inf_{(K+1)(s+1)\le \delta_0n/\log p} P_{\bSigma}\{\bX\in \scrX_{s,n,p}({c}_*,\delta_1,\xi,K) \}\ge 1- e^{-\delta_2n}. 
\ees
Consequently, when the $\bX$ in (\ref{LM}) is indeed generated from (\ref{P-measure}) and (\ref{random-X}), 
all conclusions of Theorems \ref{th-1}, \ref{cor-1}, and \ref{cor-2} hold 
for both (\ref{scaled-Lasso}) and (\ref{lse-after}) with an adjustment of a probability smaller than 
$2e^{-\delta_2n}$, 
provided that $\bbeta\in \scrB_1(s,\sigma\lam_{univ})$ and $\lam_0=A\sqrt{(2/n)\log(p/\eps)}$ 
in (\ref{scaled-Lasso}) with a fixed $A> (\xi+1)/(\xi-1)$. 
\end{theorem}

\begin{remark}\label{remark-th3} 
It follows from Theorem II.13 of \cite{DavidsonS01} that 
for certain positive $\{\delta_0,\delta_1,\delta_2\}$, 
\bes
\scrX'_{n,p} = \Big\{\bX: \min_{|S|+m\le \delta_0n/\log p}\phi_-(m,S)\ge {c}_*(1-\delta_1), 
\max_{|S|+m\le \delta_0 n/\log p}\phi_+(m,S)\le c^*(1+\delta_1)\Big\}
\ees
satisfies $P_{\bSigma}\{\scrX_{n,p}'\}\ge 1- e^{-\delta_2 n}$ for all $\bSigma$ in Theorem \ref{th-3}
\cite{CandesT05,ZhangH08}. 
Let $K=4\xi^2(c^*/{c}_*)(1+\delta_1)/(1-\delta_1)$ and $\{k,\ell\}$ be positive integers satisfying 
$4\ell/k\ge K$ and $\max\{k + \ell,4\ell\}\le \delta_0 n/\log p$. 
For $\bX\in \scrX'_{n,p}$, the conditions 
\bes
\kappa(\xi,S)\ge \{{c}_*(1-\delta_1)\}^{1/2}/2,\ 
\xi^2\phi_+(m,S)/\kappa^2(\xi,S)\le K,
\ees
hold for all $|S|\le k$, where $m$ is smallest integer upper bound of $K|S|$.  
\end{remark}

The $P_{\bSigma}$-induced regression model (\ref{LM-j}) provides a motivation for the use of  the Lasso 
in (\ref{Lasso-path-j}) and Table~\ref{table:alg} to generate score vectors $\bz_j$. 
However, the goal of the procedure is to find $\bz_j$ with small $\eta_j$ and $\tau_j$ for controlling 
the variance and bias of the LDPE (\ref{LDPE}) as in Theorem~\ref{th-1}. 
This is quite different from the usual applications of the Lasso for prediction, estimation 
of regression coefficients, or model selection. 


\section{Simulation Results}
We set $n=200$, $p=3000$, and run several simulation experiments with 
100 replications in each setting. 
In each replication, we generate an independent copy of $(\tbX,\bX,\by)$, 
where, given a particular $\rho \in(-1,1)$, $\tbX = (\xtil_{ij})_{n\times p}$ 
has iid $N(0,\bSigma)$ rows with $\bSigma = (\rho^{|j-k|})_{p\times p}$, 
$\bx_j = \tbx_j\sqrt{n}/{|}\tbx_j{|}_2$, and $(\bX,\by)$ is as in (\ref{LM}) with $\sigma=1$. 
Given a particular $\alpha \geq 1$, $\beta_j = 3\lam_{univ}$ for $j = 1500, 1800, 2100, \ldots, 3000$, 
and $\beta_j = 3\lam_{univ}/j^\alpha$ for all other $j$, where $\lam_{univ}=\sqrt{(2/n)\log p}$. 
Our simulation design is set to test the performance of the LDPE methods 
beyond the assumptions of the theorems in Section~3;
this setup gives 
$(s,s*(\log p)/n^{1/2})=(8.93,5.05)$ and $(29.24,16.55)$ respectively for 
$\alpha=2$ and $1$, while the theorems require $s(\log p)/\sqrt{n}\to 0$, 
where $s=\sum_j\min(|\beta_j|/\lam_{univ},1)$. 
This simulation example includes four cases, labeled (A), (B), (C), and (D), respectively:
$(\alpha,\rho)=(2,1/5)$, $(1,1/5)$, $(2,4/5)$, and $(1,4/5)$, with case (D) being the most difficult one. 

\begin{centering}
\begin{figure}[ht]
\includegraphics{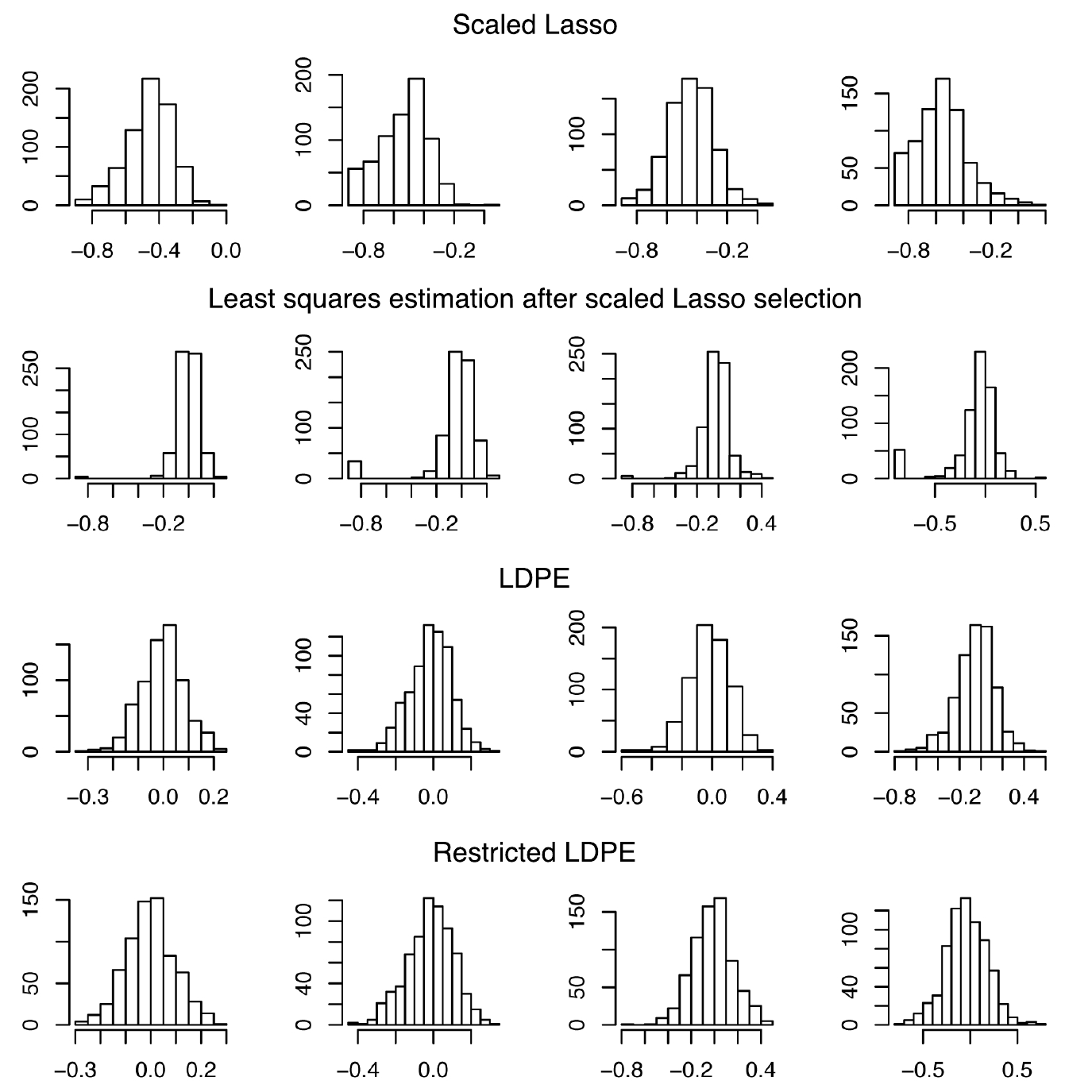}
\caption{Histogram of errors when estimating maximal $\beta_j$ using the scaled Lasso, 
{the scaled Lasso-LSE}, the LDPE, and the {R-LDPE}. From left to right, 
plots correspond to simulation settings (A), (B), (C), and (D).}
\label{fig:hist}
\end{figure}
\end{centering}

\begin{table}[ht]
\begin{tabular}{rlcccccc}
\toprule
&&\multicolumn{6}{c}{Estimator}\\
\cmidrule{3-8}
&& Lasso & scLasso & {scLasso-LSE}& oracle & LDPE & R-LDPE\\ 
\midrule
\addlinespace
(A) & bias & -0.2965 & -0.4605 & -0.0064 & -0.0045 & -0.0038 & -0.0028\\
 & sd& 0.0936 & 0.1360 & 0.1004 & 0.0730 & 0.0860 & 0.0960\\
 & median abs error& 0.2948 & 0.4519 & 0.0549 & 0.0507 & 0.0531 & 0.0627\\
\addlinespace
(B) & bias & -0.2998 & -0.5341 & -0.0476 & 0.0049 & -0.0160 & -0.0167\\
 &sd & 0.1082 & 0.1590 & 0.2032 & 0.0722 & 0.1111 & 0.1213\\
 & median abs error & 0.2994 & 0.5150 & 0.0693 & 0.0500 & 0.0705 & 0.0799\\
\addlinespace
(C) &bias & -0.3007 & -0.4423 & -0.0266 & -0.0049 & -0.0194 & -0.0181\\
 & sd & 0.1207 & 0.1520 & 0.1338 & 0.1485 & 0.1358 & 0.1750\\
 & median abs error & 0.3000 & 0.4356 & 0.0657 & 0.0994 & 0.0902 & 0.1150\\
\addlinespace
(D)& bias& -0.3258 & -0.5548 & -0.1074 & -0.0007 & -0.0510 & -0.0405\\
&sd& 0.1367 & 0.1844 & 0.2442 & 0.1455 & 0.1768 & 0.2198\\
 &median abs error& 0.3319 & 0.5620 & 0.0857 & 0.0955 & 0.1112 & 0.1411\\
\bottomrule
\addlinespace
  \end{tabular}
\caption{Summary statistics for various estimates of the maximal $\beta_j = {|}\bbeta{|}_\infty$: the Lasso, the scaled Lasso (scLasso), {the scaled Lasso-LSE} (scLasso-LSE), the oracle estimator, the LDPE, and the {R-LDPE}.}
\label{table:maxerror}
\end{table}

In addition to the Lasso with penalty level $\lam_{univ}$, the scaled 
Lasso (\ref{scaled-Lasso}) with penalty level $\lam_0=\lam_{univ}$, 
and {the scaled Lasso-LSE} (\ref{lse-after}), 
we consider an oracle estimator along with the LDPE  (\ref{LDPE})
and its restricted version derived from (\ref{z_j-2nd}), {the} R-LDPE. 
The oracle estimator is the the least squares estimator of $\beta_j$ 
when {the} $\beta_k$ are given for all $k\neq j$ except for those $k$ with $|k-j|$ 
among the smallest three. It can be written as 
\bel{oracle}
\hbeta_j^{(o)} 
= \frac{(\bz_j^{(o)})^T}{\|\bz_j^{(o)}\|_2^2}\Big(\by - \sum_{k\not\in K_j} \bx_k\beta_k\Big),\ 
\hsigma^{(o)} = \|\bP_{K_j}^\perp\bep\|_2/\sqrt{n},
\eel
where $K_j=\{j-1,j,j+1\}$ for $1<j<p$, $K_1=\{1,2,3\}$, $K_p=\{p-2,p-1,p\}$, {and}
$\bz_j^{(o)} = \bP_{K_j\setminus\{j\}}^\perp\bx_j$. {Here},
$\bP_{K}^\perp$ is the orthogonal projection to the space of $n$-vectors orthogonal to $\{\bx_k, k\in K\}$. 
Note that the oracular knowledge reduces the complexity of the problem from $(n,p)=(200,3000)$ to $(n,p)=(200,3)$, 
and that the variables $\{\bx_k, k\in K_j\}$ also have the highest correlation to $\bx_j$. 
For both the LDPE and the {R-LDPE}, 
{the scaled Lasso-LSE} (\ref{lse-after}) is used to generate $\hbbeta^{(init)}$ and $\hsigma$,
while the algorithm in Table \ref{table:alg} is used to generate $\bz_j$, with $\kappa_0=1/4$. 
The default $\eta^*_j=\sqrt{2\log p}$ passed the test in Step 1 of Table \ref{table:alg} without adjustment 
in all instances in the simulation study. This guarantees $\eta_j\le \sqrt{2\log p}$ for the bias factor. 
For the {R-LDPE}, $m=4$ is used in (\ref{z_j-2nd}).

The asymptotic normality of the LDPE holds well in our simulation experiments.  
Table ~\ref{table:maxerror} and Figure ~\ref{fig:hist} demonstrate the behavior of the LDPE and {R-LDPE} for 
the largest $\beta_j$, compared with that of the other estimation methods. 
The scaled Lasso has more bias and a larger variance than the Lasso, but is entirely data-driven. 
The bias can be significantly reduced though {the scaled Lasso-LSE}; 
however, error resulting from failure to select some maximal $\beta_j$ remains. 
This is clearest in the histograms corresponding the distribution of errors for {the scaled Lasso-LSE} 
in settings (B) and (D), where $\alpha = 1$ and the $\beta_j$ decay at a slower rate. 
For a small increase in variance, the LDPE and {R-LDPE} further reduce the bias of the {scaled Lasso-LSE}. 
This is also the case 
when $\hbbeta^{(init)}$ is a heavily biased estimator such as the Lasso or scaled Lasso, 
and the  improvement is most dramatic when estimating large $\beta_j$. 
Although the asymptotic normality of the LDPE holds even better for small $\beta_j$ in the simulation study, 
a parallel comparison for small $\beta_j$ is not meaningful; the Lasso typically estimates small $\beta_j$ by zero, 
while the raw LDPE is not designed to be sparse. 

\begin{table}[ht]
\begin{tabular}{llcccc}
\toprule
&&(A)&(B)&(C)&(D)\\
\midrule
all $\beta_j$& LDPE & 0.9597 & 0.9845 & 0.9556 & 0.9855 \\
 &R-LDPE & 0.9595 & 0.9848 & 0.9557 & 0.9885  \\
 maximal $\beta_j$ & LDPE&  0.9571 & 0.9814& 0.9029 &0.9443\\
& R-LDPE & 0.9614 & 0.9786 & 0.9414 & 0.9786\\
\bottomrule
\end{tabular}
\caption{Mean coverage probability of LDPE and R-LDPE.}
\label{table:meancover}
\end{table}

\begin{centering}
\begin{figure}
\includegraphics{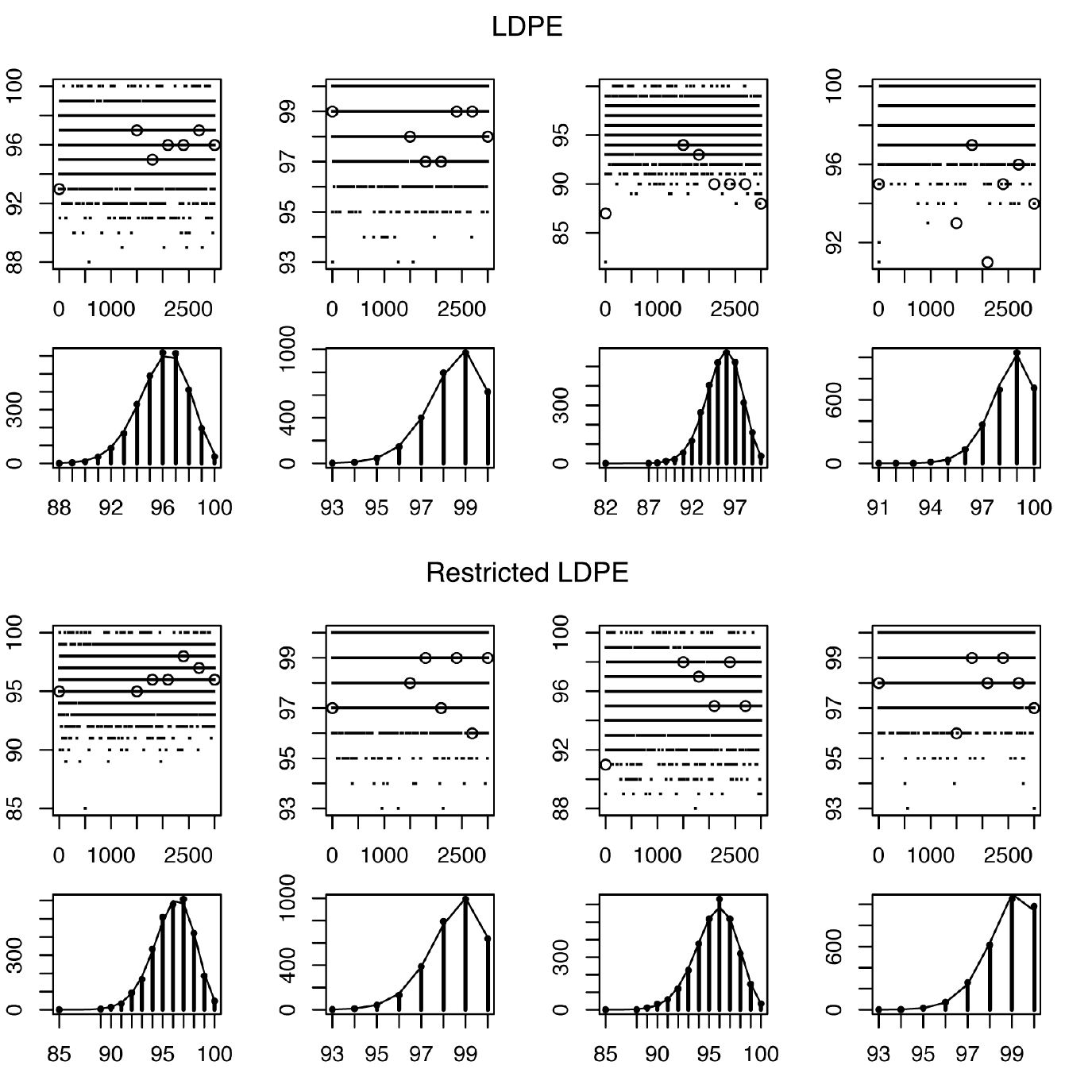}
\caption{Rows 1 and 3: Coverage frequencies versus the index of $\beta_j$. 
Points corresponding to maximal $\beta_j$ are plotted as large circles. 
Rows 2 and 4: The number 
of variables for given values of the relative coverage frequency, 
superimposed {on} the {binomial$(100,\tilde p)$} probability mass function, 
{where $\tilde p$ is the simulated mean coverage}. 
Figures depict results from simulations (A), (B), (C), and (D), from left to right.}
\label{fig:cover}
\end{figure}
\end{centering}

The overall coverage probability of the LDPE-based confidence interval matches relatively 
well to the preassigned level, as expected from our theoretical results. 
The LDPE and R-LDPE create confidence intervals $\hbeta_j\pm 1.96\hsigma \tau_j$ with 
approximately 95\% coverage in settings (A) and ({C}) and somewhat higher coverage probability in (B) and (D). 
Refer to Table ~\ref{table:meancover} for precise values. 
Since the coverage probabilities for each individual $\beta_j$ are calculated based on a sample of 
100 replications, the empirical distribution of the simulated relative coverage frequencies 
exhibits some randomness, which matches that of the 
binomial$(n,\ptil)$ distribution, with $n = 100$ and $\ptil$
equal to the simulated mean coverage, as shown in Figure~\ref{fig:cover}. 

Two separate issues may lead to some variability in the coverage. 
As is the case with settings (B) and (D), overall coverage may exceed the stated confidence level 
when presence of many small signals in $\beta$ is interpreted as noise, increasing $\hsigma$ 
and hence the width of the confidence intervals, along with the coverage; 
however, this phenomenon will not result in under-coverage.
In addition, compared with the overall coverage probability, the coverage probability is somewhat smaller 
when large values of $\beta_j$ are associated with highly correlated columns of $\bX$. 
This is most apparent when plotting coverage versus index in (C) and (D), 
the two settings with higher correlation between adjacent columns of $\bX$.
For additional clarity, the points corresponding to maximal values of 
$\beta_j$ in Figure~\ref{fig:cover} are emphasized by larger circles, and the coverage of the LDPE and R-LDPE for maximal $\beta_j$ are listed separately from the overall coverage in the last two rows of 
Table~\ref{table:meancover}. 
It can be seen from these details
that the R-LDPE (\ref{z_j-2nd}) further eliminates 
the bias caused by relatively large values of $\beta_j$ associated with highly correlated columns of $\bX$ 
and improves coverage probabilities. 
The bias correction effect can be also seen in the histograms in Figure \ref{fig:hist} in setting (D), 
but not in (C). 

\begin{figure}
\includegraphics{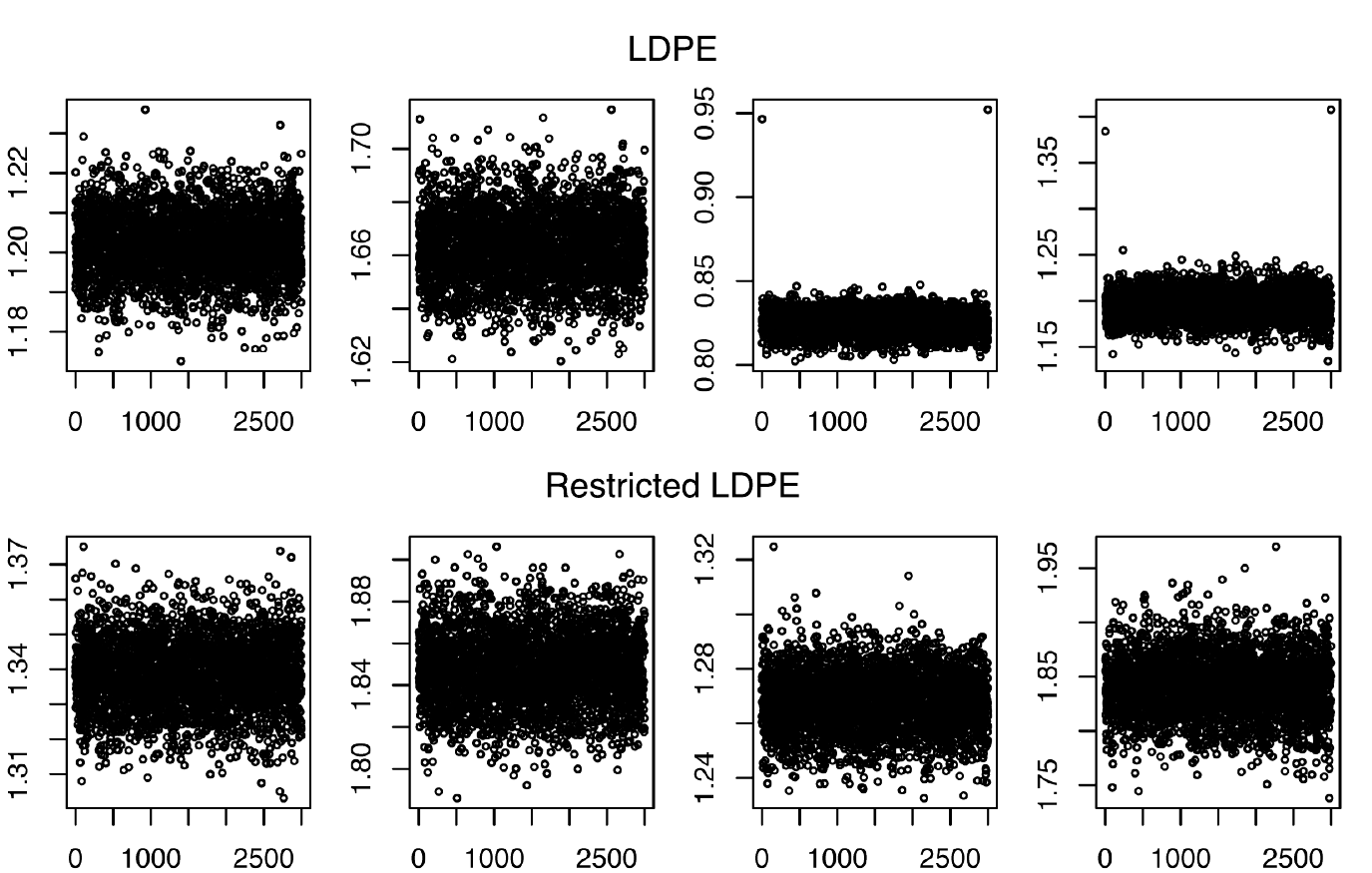}
\caption{Median ratio of width of the LDPE and R-LDPE confidence intervals versus the oracle confidence interval for each $\beta_j$.}
\label{fig:width}
\end{figure}

\begin{table}[ht]
\begin{tabular}{lrrrr}
\toprule
& (A) & (B) & (C) & (D) \\
\midrule
LDPE & 1.2020 & 1.6400 & 0.8209 & 1.1758 \\
R-LDPE & 1.3359 & 1.8238 & 1.2678 & 1.8150 \\
\bottomrule
\end{tabular}
\caption{Median of the width ratio medians in Figure \ref{fig:width}.}
\label{tab:width}
\end{table}

The LDPE and R-LDPE confidence intervals are of reasonable width, 
comparable to that of the confidence intervals derived from the oracle estimator. 
Consider the median ratio between the width of the LDPE  (and restrictd LDPE) confidence intervals 
and the oracle confidence intervals, shown in Figure~\ref{fig:width}. 
The distribution of the median ratio associated with each $\beta_j$ is uniform 
over the different $j = 1, \ldots, 3000$ in settings (A) and (B). 
The anomalies at $j = 1$ and $j = 3000$ in settings (C) and (D) are a result of the structure of $\bX$. 
When the correlation between nearby columns of $\bX$ is high, the fact that the first and last columns of $\bX$ 
have fewer highly-correlated neighbors gives the oracle {a} relatively {greater} advantage. 
Since the medians of the ratios are uniformly distributed over $j$, it is reasonable to summarize the ratios in each simulation setting with the median value over every replication of every $\beta_j$, as listed in Table~\ref{tab:width}. 
Note that the LDPE is more efficient than the oracle estimator in the high-correlation settings (C) and (D). 
This is probably due to the benefit of relaxing the orthogonality constraint of $\bx^\perp_j$  
when the correlation of the design is high and the error of the initial estimator is relatively small. 
The median ratio between the widths for the LDPE estimator reaches its highest value of 1.6400 in setting (B), 
where the coverage of the LDPE intervals is high and the benefit of relaxing the orthogonality constraint is small, 
if any, relative to the oracle. 

Recall that the R-LDPE improves the coverage probability for large $\beta_j$ at the cost of an 
increase in the variance of the estimator; thus, the R-LDPE confidence intervals are somewhat
wider than the LDPE confidence intervals. Although the improvement in coverage probability is focused 
on the larger values of $\beta_j$, all $\beta_j$ are affected by the increase in variance and confidence interval width.

\begin{centering}
\begin{figure}[ht]
\includegraphics{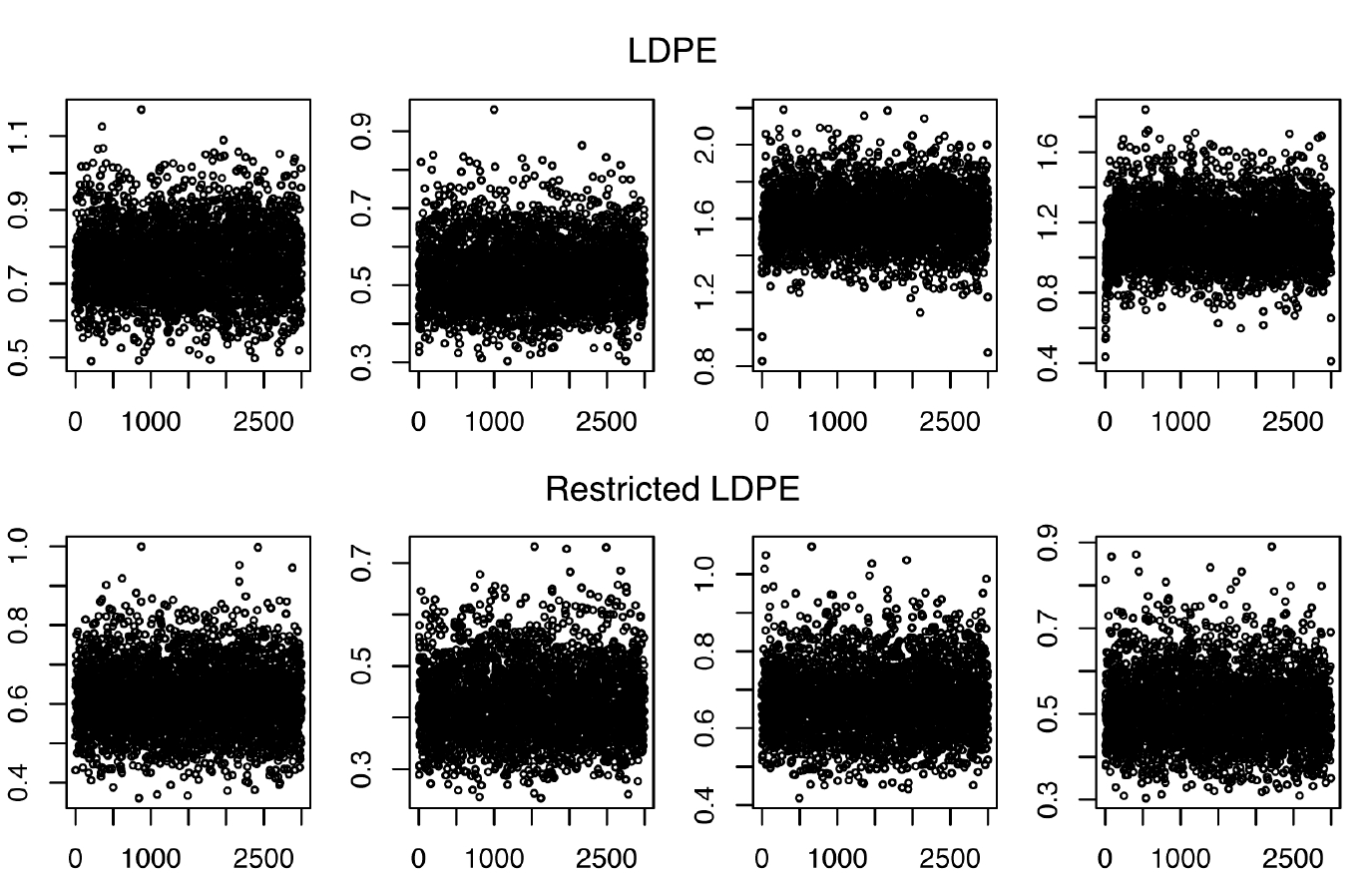}
\caption{Efficiency (the ratio of the MSE's) of the LDPE and R-LDPE estimators versus the oracle estimator for each $\beta_j$.} 
\label{fig:eff}
\end{figure}
\end{centering}
\begin{table}[ht]
\begin{tabular}{lrrrr}
\toprule
& (A) & (B) & (C) & (D) \\
\midrule
LDPE & 0.7551 & 0.5232 & 1.5950 & 1.1169 \\
R-LDPE & 0.6086 & 0.4232 & 0.6656 & 0.5049 \\
\bottomrule
\end{tabular}
\caption{Medians of the MSE ratios in Figure \ref{fig:eff}.}
\label{tab:eff}
\end{table}

We may also consider the performance of LDPE as a point estimator. 
Table~\ref{tab:eff} and Figure ~\ref{fig:eff} compare the MSEs of the LDPE and R-LDPE 
estimators $\beta_j$ to that of the oracle estimator of $\beta_j$. 
This comparison is consistent with the comparison of the median width of confidence intervals 
in Table \ref{tab:width} and Figure \ref{fig:width} discussed earlier.

The Lasso and scaled Lasso estimators have larger biases for bigger values of $\beta_j$ 
but perform very well for smaller values. 
On the other hand, the LDPE and the oracle estimator are not designed to be sparse 
and has very stable errors over the $\beta_j$. 
For the estimation of the entire vector $\bbeta$ or its support, it is appropriate to compare a thresholded 
LDPE with the Lasso, the scaled Lasso, the {scaled Lasso-LSE}, and a matching thresholded oracle estimator. 
Hard thresholding was implemented: 
$\hbeta_j I\{|\hbeta_j|\le \that_j\}$ for the thresholded LDPE with $\that_j = \hsigma\tau_j\Phi^{-1}(1-1/(2p))$ 
and $\hbeta_j^{(o)} I\{|\hbeta_j^{(o)}|\le \that_j^{(o)}\}$ for the thresholded oracle with 
$\that_j^{(o)} = \hsigma^{(o)}\|\bz_j^{(o)}\|_2^{-1}\Phi^{-1}(1-1/(2p))$, where 
$\{\hbeta_j^{(o)},\hsigma^{(o)},\bz_j^{(o)}\}$ are as in (\ref{oracle}). 
Since $\beta_j\neq 0$ for all $j$, the comparison is confined to the $\ell_2$ estimation error. 
Table ~\ref{table:L2} lists the mean, standard deviation, and median of 
the $\ell_2$ loss of these five estimators over 100 replications.  
Of the five estimators, only the scaled Lasso, the {scaled Lasso-LSE}, and the thresholded LDPE are purely data-driven. 
The performance of the {scaled Lasso-LSE}, thresholded LDPE, and thresholded oracle are comparable 
and they always outperform the scaled Lasso. 
They also outperform the Lasso in cases (A), (B), and (C). 
In the hardest case, (D), which has both a high correlation between adjacent columns of $\bX$ 
and a slower decay in $\beta_j$, the thresholded oracle slightly outperforms the Lasso and 
the Lasso sightly outperforms the {scaled Lasso-LSE} and thresholded LDPE.
Generally, the $\ell_2$ loss of the thresholded LDPE remains 
slightly above that of the {scaled Lasso-LSE}, which improves upon the scaled Lasso 
by reducing its bias. 
Note that our goal is not to find a better estimator for the entire vector $\bbeta$ since 
quite a few versions of estimation optimality of regularized estimators have already been established. 
What we demonstrate here is that the cost of removing the bias with the LDPE, and thus giving up shrinkage, 
is small.

\begin{table}
\begin{tabular}{rlccccc}
\toprule
&&\multicolumn{5}{c}{Estimator}\\
\cmidrule(lr){3-7}
&& Lasso & scLasso &{scaled Lasso-LSE} & T-oracle & T-LDPE \\ 
\midrule
 \addlinespace
(A) & mean & 0.8470 & 1.2706 & 0.3288 & 0.3624 & 0.3621\\
 & sd &0.1076 & 0.2393 & 0.1465 & 0.0908 & 0.1884\\
 & median & 0.8252&1.2131 & 0.3042 & 0.3577 & 0.3312\\
 \addlinespace
(B) & mean& 0.9937& 1.5837 & 0.7586 & 0.5658 & 0.7969\\
 & sd &0.1214& 0.2624 & 0.2976 & 0.0615 & 0.3873\\
& median &0.9820& 1.5560 & 0.6219 & 0.5675 & 0.6983\\
 \addlinespace
(C) & mean &0.8836& 1.2411 & 0.4817 & 0.6803 & 0.5337\\
 & sd &0.1402& 0.2208 & 0.2083 & 0.2843 & 0.2164\\
 & median &0.8702& 1.2295 & 0.4343 & 0.6338 & 0.4642\\
 \addlinespace
(D) & mean &1.0775& 1.6303 & 1.0102 & 0.9274 & 1.2627\\
 & sd & 0.1437&0.2381 & 0.3572 & 0.2342 & 0.5576\\
 & median &1.0570& 1.6389 & 0.9216 & 0.8716 & 1.1011\\
\bottomrule
\addlinespace
\end{tabular}
\caption{Summary statistics for the $\ell_2$ loss of five estimators of $\bbeta$: 
the Lasso, the scaled Lasso, the {scaled Lasso-LSE}, the thresholded oracle estimator (T-oracle), 
and the thresholded LDPE (T-LDPE)}.
\label{table:L2}
\end{table}

\section{Discussion}

We have developed the LDPE method of constructing $\hbeta_1,\ldots,\hbeta_p$ 
for the individual regression coefficients and estimators for their finite dimensional covariance structure.  
Under proper conditions on $\bX$ and $\bbeta$, we have proven the asymptotic unbiasedness and normality 
of the finite-dimensional distribution functions of these estimators and the consistency 
of their estimated covariances. 
Thus, LDPE yields an approximate Gaussian sequence as in (\ref{Gaussian-seq}), also called raw LDPE, 
which allows one to assess the level of significance of each unknown coefficient $\beta_j$ 
without the uniform signal strength assumption (\ref{unif-sig}), compared with the existing 
variable selection approach. 
The proposed method applies to making inference about a preconceived low-dimensional parameter, 
an interesting practical problem and a primary goal of this paper.  
It also applies to making inference about all regression coefficients via simultaneous interval estimation 
and correct selection of large and zero coefficients in the presence of many small coefficients. 

The raw LDPE estimator is not sparse, but it can be thresholded to take advantage of the sparsity 
of $\bbeta$, and the sampling distribution of the thresholded LDPE can still be bounded based on 
the approximate distribution of the raw LDPE. A thresholded LDPE is proven to attain $\ell_2$ rate 
optimality for the estimation of an entire sparse $\bbeta$. 

The focus of this paper is interval estimation and hypothesis testing without the uniform 
signal strength condition. 
Another important problem is prediction. 
Since prediction at a design point $\ba$ is equivalent to the estimation of 
the ``contrast'' $\ba^T\bbeta$, with possibly large $\|\ba\|_0$, the implication of 
LDPE on prediction is an interesting future research direction. 

We use the Lasso to provide a relaxation of the projection of $\bx_j$ to 
$\bx_j^\perp$. This choice is primarily due to our familiarity with the computation of the Lasso  
and the readily available scaled Lasso method of choosing a penalty level. 
We have also considered some other methods of relaxing the projection. 
Among these other methods, a particularly interesting one is the following constrained 
minimization of the variance of the noise term in (\ref{error-decomp}): 
\bel{q-prog}
\bz_j = \argmin_{\bz}\Big\{\|\bz\|_2^2: |\bz_j^T\bx_j| = n, 
\max_{k\neq j}|\bz_j^T\bx_k/n|\le\lam_j'\Big\}. 
\eel
Similar to the Lasso in (\ref{Lasso-z_j}), (\ref{q-prog}) is a quadratic programme. 
The Lasso solution (\ref{Lasso-z_j}) is feasible in (\ref{q-prog}) with 
$\lam_jn/|\bz_j^T\bx_j| = \lam_j'$. Our results on these and other extensions of 
our ideas and methods will be presented in a forthcoming paper. 

\section{Appendix} 

{\bf Proof of Proposition \ref{prop-1}.} (i) For $\hbgamma_j(\lam)=0$, $\|\bz_j(\lam)\|_2=\|\bx_j\|_2=\sqrt{n}$ 
and $\eta_j(\lam)=\max_{k\neq j}|\bx_k^T\bx_j|/\sqrt{n}$ do not depend on $\lam$. 
Consider $\hbgamma_j(\lam)\neq 0$. 
Since $\hbgamma_j(\lam)$ is continuous and piecewise linear in $\lam$, it suffices 
to consider a fixed open interval $\lam\in I_0$ in which $\bs = \sgn(\hbgamma_j(\lam))$ do not change with $\lam$. 
Let $A = \{k\neq j: s_k\neq 0\}$, and let $\bQ_A$ be the projection operator $\bb \to \bb_A$. 
It follows from the Karush-Kuhn-Tucker conditions for the Lasso that 
\bes
\bX_A^T\bz_j(\lam) = 
\bX_A^T\{\bx_j-\bX_A\bQ_A\hbgamma_j(\lam)\}/n = \bX_A^T\{\bx_j-\bX_{-j}\hbgamma_j(\lam)\}/n = \lam \bs_A. 
\ees
This gives 
$(\pa/\pa\lam)\bQ_A\hbgamma_j(\lam) = - (\bX_A^T\bX_A/n)^{-1}\bs_A$ for all $\lam\in I_0$. 
It follows that 
\bes
(\pa/\pa\lam)\|\bz_j(\lambda)\|_2^2 
&=&(\pa/\pa\lam)\|\bx_j-\bX_A\bQ_A\hbgamma_j(\lam)\|_2^2 
\cr &=& - 2\{(\pa/\pa\lam)\bQ_A\hbgamma_j(\lam)\}^T\bX_A^T(\bx_j-\bX_A\bQ_A\hbgamma_j(\lam))
\cr &=& 2\{(\bX_A^T\bX_A/n)^{-1}\bs_A\}^T\bX_A^T\bz_j(\lambda)
= (2/\lam)\|\bP_A\bz_j(\lam)\|_2^2,
\ees
where $\bP_A=\bX_A(\bX_A^T\bX_A)^{-1}\bX_A^T$ is the projection to the column 
space of $\bX_A$. 
Thus, $\|\bz_j(\lam)\|_2$ is nondecreasing in $\lam$. 
Since  $\|\bs\|_\infty =1$, $\eta_j(\lam)=n\lam/\|\bz_j(\lam)\|_2$, so that 
\bes
(\lam^3/2) (\pa/\pa\lam)\big\{\eta_j(\lam)/n\big\}^{-2}
= (\lam^3/2) (\pa/\pa\lam)\big\{\lam^{-2}\|\bz_j(\lam)\|_2^2\big\}
= \|\bP_A\bz_j(\lambda)\|_2^2 - \|\bz_j(\lam)\|_2^2 \le 0. 
\ees
Thus, $\eta_j(\lam)$ is nondecreasing in $\lam$. 
Since $\hsigma_j(\lam)$ is the solution of $\|\bz_j(\lam\sigma)\|_2=\sigma\sqrt{n}$, 
it is also a solution of $\eta_j(\sigma\lam) = \lam\sqrt{n}$. 
For $\sigma < \hsigma_j(\lam)$, $\eta_j(\sigma\lam) \le \lam\sqrt{n}$, so 
$\|\bz_j(\sigma\lam)\|_2 \ge \sigma\sqrt{n}$. 
Thus, since smaller $\lam$ gives smaller $\|\bz_j(\lam\sigma)\|_2$, $\hsigma_j(\lam)$ is also nondecreasing in $\lam$. 
Since $\bx_j^T\bz_j(\lam)=\|\bz_j(\lam)\|_2^2+\{\bX_{-j}\hbgamma_j(\lam)\}^T\bz_j(\lam) 
= \|\bz_j(\lam)\|_2^2+\lam\|\hbgamma_j(\lam)\|_1$, we also have $\tau_j(\lam)\le 1/\|\bz_j(\lam)\|_2$.  

(ii) Since the Lasso path $\hbgamma_j(\lam)$ is continuos in $\lam$ and $\eta_j(\lam)$ is nondecreasing, 
the range of $\eta_j(\lam)$ is  an interval. Within the interior of this interval, $\hbgamma_j(\lam)\neq 0$ and 
$\eta_j(\lam) = n\lam/\|\bz_j(\lam)\|_2$. 
We have shown in the proof of (i) that $\hsigma_j(t)$ is a solution of $\eta_j(\sigma t) = t\sqrt{n}$ 
when $t\sqrt{n}$ is in the range of $\eta_j(\lam)$. 
If $\eta_j(\lam)<t\sqrt{n}$ for all $\lam$, then $\hsigma_j(t)=\|\bx_j\|_2/\sqrt{n}=1$ is attained at $\hbgamma_j(\infty)=0$. 
If $\eta_j(\lam)>t\sqrt{n}$ for all $\lam$, then $\hsigma_j(t)=0$. 
This gives (\ref{prop-1-1}). The upper bounds follow for $\eta_j^*$ and $\tau_j$. 

It remains to verify the last assertion of part (ii) for $\bz_j(0)=0$. 
Consider vectors $\bb$ with $\|\bb-\hbgamma_j(0+)\|_1\le \eps$ and the loss function in 
(\ref{scaled-Lasso-j}) at $\{t\bb,\sigma\}$ with $0\le t\le 1$. 
Since $\bX_{-j}\hbgamma(0+)=\bx_j$ and $\|\bx_j\|_2=\sqrt{n}$, 
the minimum of the loss function over $\sigma$ is approximately 
\bes
\min_{\sigma}\big\{\|\bx_j- t\bX_{-j}\bb\|_2^2/(2n\sigma)+\sigma/2 + t\lam\|\bb\|_1\big\}
=\big\{1-t+O(\eps)\big\} + t\lam\big\{\|\hbgamma(0+)\|_1+O(\eps)\big\}. 
\ees
When $\lam\|\hbgamma_j(0+)\|_1 >1$, the minimum of the above expression is 
attained at $t\approx 0$ for sufficiently small $\eps$. This gives $\hsigma_j(\lam)>0$. 
Conversely, when $\lam\|\hbgamma_j(0+)\|_1<1$, the optimal $t$ for 
$\hbgamma(\lam)$ with very small $\lam$ is $t \approx 1$, so that by the joint convexity of the loss function, $\hsigma_j(\lam)=0$.  
Since,  by (\ref{prop-1-1}), $\eta_j(0+)=\inf\{\lam\sqrt{n}:\hsigma_j(\lam)>0\}$, 
the relationship between $\eta_j(0+)$ and the $\ell_1$ minimization problem follows. 

(iii) Let $\bgamma_j$ be the solution of $\bx_j=\bX_{-j}\bgamma_j$ with the shortest $\|\bgamma_j\|_1$, 
and $\lam = 1/\|\bgamma_j\|_1$. Let 
$\bbeta_{-j}=s\lam_{univ}\lam \bgamma_j$, and 
$\beta_j=-s\lam_{univ}\lam$. Then, $\bX\bbeta=0$ and $\sum_{j=1}^p\min(|\beta_j|/\lam_{univ},1)\le s+1$. 
It follows that for the optimal $\bdelta$, 
\bes
4 C_0s\lam_{univ}^2\ge 2\|\bbeta-\bdelta\|_2^2+2\|\bdelta\|_2^2 \ge \|\bbeta\|_2^2
\ge |\beta_j|^2 = (s\lam_{univ}\lam)^2. 
\ees
Taking $s=a_0n/(\log p)$ gives $\lam^2\le (4C_0/a_0)(\log p)/n$.  
Thus, by part (ii), $\max_j \eta_j^2(0+) \le \lam^2 n \le (4C_0/a_0)\log p$. 
This implies the upper bound for $\max_{j\le p}\eta_j^*$ by Step 1. 
$\hfill\square$

\medskip
{\bf Proof of Theorem \ref{th-1}.}  The error decomposition in (\ref{error-decomp}) and (\ref{bias-sum}) implies 
\bes
\Big|\tau_j^{-1}(\hbeta_j-\beta_j) - \bz_j^T\bep/\|\bz_j\|_2\Big| 
\le \Big(\max_{k\neq j}|\bz_j^T\bx_k|/\|\bz_j\|_2\Big)\|\hbbeta^{(init)}-\bbeta\|_1
= \eta_j \|\hbbeta^{(init)}-\bbeta\|_1. 
\ees
This and (\ref{ell_1-err-bd}) yield (\ref{th-1-1}). 
When $\big|\tau_j^{-1}(\hbeta_j - \beta_j) - \bz_j^T\bep/\|\bz_j\|_2\big| 
\le \sigma^* \eps_n'$ and $|\hsigma/\sigma^* - 1 | \le \eps_n''$,  
$\tau_j^{-1}|\hbeta_j - \beta_j| \ge \hsigma t$ implies 
$|\bz_j^T\bep|/\|\bz_j\|_2\ge \hsigma t - \sigma^* \eps_n'
\ge \sigma^*\{(1-\eps_n'')t-\eps_n'\}$. 
Since $\bep\sim N(0,\sigma^2\bI)$ and $\bz_j$ depends on $\bX$ only, 
$\bz_j^T\bep/(\|\bz_j\|_2\sigma^*) \sim \sqrt{n}\veps_1/\|\bep\|_2$. Thus, for $x\ge 1$, 
\bes
P\big\{|\bz_j^T\bep|/\|\bz_j\|_2 \ge \sigma^*x \big\}
= P\big\{ (n-x^2)\veps_1^2 \ge x^2(\veps_2^2+\cdots+\veps_n^2)\} \le 2\Phi_n(-x). 
\ees 
The same argument also implies (\ref{th-1-3}) with fixed $m$, 
since $\max(\eps_n',\eps_n'')\to 0$ and $\bV$ in (\ref{cov-LDPE}) is the approximate 
covariance between $\hbeta_j$ and $\hbeta_k$. $\hfill\square$

\medskip
{\bf Proof of Theorem \ref{cor-1}.} Since (\ref{th-1-1}) is uniform in $\eps\in [\alpha_0/p^2,1]$, 
(\ref{cor-1-1}) and (\ref{cor-1-2}) follow directly. By Lemma 1 of \cite{SunZ11}, 
$2\Phi_n(-\sqrt{n\{\exp(2t^2/(n-1))-1\}})\le (\pi^{-1/2}+o(1))e^{-t^2}/t$ as 
$\min(n,t)\to\infty$. When $s\log p=o(n^{1/2})$, $\eps_n'=o(1)$ and $\eps_n''=o(n^{-1/2})$. 
Let $t = \sqrt{2\log(p/\alpha)}+c_0$. Since $\log(p/\alpha) \ll n$ and $\alpha$ is fixed, 
\bes
 - (1-\eps_n'')t + \eps_n' = - t+o(1) = - \sqrt{n\{\exp(2t^2/(n-1))-1\}}-c_0+o(1). 
\ees
Thus, the right-hand side of (\ref{cor-1-2}) is no greater than $\alpha$ in the limit. 
This and a similar inequality with the true $\sigma$ yields (\ref{cor-1-3}). $\hfill\square$ 

\medskip
The following lemma, needed in the proof of Theorem \ref{cor-2}, 
controls the loss of a perturbed soft threshold estimator. 
It extends Lemma 8.3 of \cite{Johnstone98} and Lemma 6.2 of \cite{Zhang05}. 

\begin{lemma}\label{lm-1} Let $s_t(x)=\sgn(x)(|x|-t)_+$, $z = \mu+\veps$ with $\veps \sim N(0,\sigma^2)$. 
Suppose that for certain constants $t$ and $\Delta$, 
$|\zhat - z|+|\that - t| \le \Delta\le t$ and $\that > t + |\zhat - z|$ in an event $\Omega$. Then, 
\bes
&& E\big\{s_{\that}(\zhat)-\mu\big\}^2I_{\Omega}
\cr &\le& \min\Big\{2E(\veps - t)_+^2 + \mu^2,\sigma^2+(t+\Delta)^2\Big\}
+\Delta\Big\{2E(\veps-t)_++3\Delta P(\veps>t)\Big\}
\cr &\le& \min\Big\{\mu^2,\sigma^2+(t+\Delta)^2\Big\}
+\varphi(t/\sigma)\Big\{4\sigma^5/t^3+2\Delta \sigma^3/t^2+3\Delta^2\sigma/t\Big\}, 
\ees
where $\varphi(x)$ and $\Phi(x)$ are the $N(0,1)$ density and distribution functions. 
\end{lemma}

\medskip
{\bf Proof.} Assume without loss of generality that $\mu>0$. Let 
\bes
f_{t,\Delta}(z,\mu) =  |-(z+t)_- - \mu| I_{\{z<0\}} + |(z - t - \Delta)_+ - \mu| I_{\{z>0\}}. 
\ees
By assumption $z+t \le \zhat + \that $ and $z-t-\Delta \le \zhat - \that\le z-t$. 
Since $s_t(z) = (z-t)_+ - (z+t)_-$, 
\bes
|s_{\that}(\zhat)-\mu|^2I_\Omega 
&\le& |-(z+t)_- - \mu|^2I_{\{z<0\}} + \big\{ |(z - t - \Delta)_+ - \mu|  + \Delta I_{\{ z -t > \mu\}} \big\}^2I_{\{z>0\}}
\cr &\le& f^2_{t,\Delta}(z,\mu) + \Delta\{2(z - t - \mu+\Delta)+\Delta\} I_{\{z>t+\mu\}}. 
\ees
Since $(\pa/\pa\mu) f^2_{t,\Delta}(\veps+\mu,\mu) = 2\mu I_{\{-t<\veps+\mu<t+\Delta\}}$, 
$Ef^2_{t,\Delta}(\veps+\mu,\mu)\le Ef^2_{t,\Delta}(\veps,0)+\int_0^\mu 2xdx$ and 
$Ef^2_{t,\Delta}(\veps+\mu,\mu)\uparrow \sigma^2+(t+\Delta)^2$. 
Since $Ef^2_{t,\Delta}(\veps,0)\le Ef^2_{t,0}(\veps,0)=2E(\veps-t)_+^2$, we have 
\bes
Ef^2_{t,\Delta}(\veps+\mu,\mu)\le \min\Big\{2E(\veps - t)_+^2 + \mu^2,\sigma^2+(t+\Delta)^2\Big\}. 
\ees
Thus, the first inequality follows from 
$E\{2(\veps - t +\Delta)+\Delta\} I_{\{\veps >t\}} = 2E(\veps-t)_+ +3\Delta P(\veps>t)$, 
and the second from $E(\veps - t)_+^k \le \sigma^k \varphi(t/\sigma)\int_0^\infty x^k e^{-xt/\sigma}dx
= k!\sigma^{2k+1}\varphi(t/\sigma)/t^{k+1}$. $\hfill\square$

\medskip
{\bf Proof of Theorem \ref{cor-2}.} 
We first prove the equivalence of the following two statements: 
\bel{pf-cor-2-1}
&&(\hsigma/\sigma)\vee(\sigma/\hsigma) -1+\eps_n'\sigma^*/
(\hsigma\wedge \sigma) \le \{1-(\hsigma/\sigma-1)_+\}{c}_n; 
\\ \label{pf-cor-2-3} && \ttil_j + \eps_n'(\sigma^*/\sigma)\ttil_j\le \that_j= (1+{c}_n)(\hsigma/\sigma)\ttil_j,\ 
\that_j-\ttil_j + \eps_n'(\sigma^*/\sigma)\ttil_j \le 2{c}_n\ttil_j. 
\eel
For $\hsigma \le \sigma$, (\ref{pf-cor-2-1}) is equivalent to 
$\sigma/\hsigma-1+\eps_n'\sigma^*/\hsigma \le {c}_n$, and (\ref{pf-cor-2-3}) to 
$\ttil_j + \eps_n'(\sigma^*/\sigma)\ttil_j \le (1+{c}_n)(\hsigma/\sigma)\ttil_j$. 
For $\hsigma > \sigma$, (\ref{pf-cor-2-1}) is equivalent to 
$\hsigma/\sigma-1+\eps_n'\sigma^*/\sigma \le (2-\hsigma/\sigma){c}_n$, and (\ref{pf-cor-2-3}) to 
$(1+{c}_n)(\hsigma/\sigma-1)\ttil_j + \eps_n'(\sigma^*/\sigma)\ttil_j \le {c}_n\ttil_j$. 
After canceling $\ttil_j$ and some algebra, we observe that 
(\ref{pf-cor-2-1}) and (\ref{pf-cor-2-3}) are equivalent in both cases. 

Let $\tveps_j = \tau_j\bz_j^T\bep/\|\bz_j\|_2 \sim N(0,\tau_j^2\sigma^2)$, 
$\tbeta_j = \beta_j + \tveps_j$, and 
\bes
\Omega_n =\big\{|\tbeta_j-\hbeta_j|\le \eps_n'(\sigma^*/\sigma)\ttil_j,\ 
\hbox{(\ref{pf-cor-2-3}) holds,}\ \forall j\le p\big\}. 
\ees
As in the proof of Theorem \ref{th-1}, 
$|\tbeta_j-\hbeta_j|\le\tau_j\eta_j\|\hbbeta^{(init)}-\bbeta\|_1$. 
Since $\max_{j\le p}\eta_jC_1s/\sqrt{n}\le\eps_n'$, 
we have $|\tbeta_j-\hbeta_j|\le \eps_n'(\sigma^*/\sigma)\ttil_j$ 
when $\|\hbbeta^{(init)}-\bbeta\|_1\le C_1s\sigma^*L_0/\sqrt{n}$. 
Thus, $P\{\Omega_n\}\ge 1-3\eps$ by (\ref{ell_1-err-bd}) and (\ref{cor-2-0}). 
Consider the event $\Omega_n$ in the rest of the proof, so that (\ref{pf-cor-2-3}) gives 
\bes
\that_j\ge\ttil_j+|\hbeta_j-\tbeta_j|,\ 
|\hbeta_j-\tbeta_j|+|\that_j-\ttil_j|\le 2{c}_n\ttil_j.
\ees
 
Since $\tveps_j \sim N(0,\tau_j^2\sigma^2)$ and $\ttil_j/(\tau_j\sigma)=L_0$, 
it follows from Lemma \ref{lm-1} with $\Delta = 2c_n\ttil_j$ that 
\bes
E\|\hbbeta^{( thr)} - \bbeta\|_2^2I_{\Omega_n}
&\le & \sum_{j=1}^p\Big[\min\Big\{\beta_j^2,\tau_j^2\sigma^2+\ttil_j^2(1+2{c}_n)^2\Big\}
\cr &&\qquad  +\varphi(L_0)\Big\{4\tau_j^2\sigma^2/L_0^3+
4{c}_n \tau_j^2\sigma^2/L_0+12{c}_n^2\tau_j^2\sigma^2L_0\Big\}\Big]. 
\ees
This gives (\ref{cor-2-1}) since $\varphi(L_0) = \eps/p$. 

Since $\that_j\ge\ttil_j+|\hbeta_j-\tbeta_j|$, 
$|\hbeta_j|>\that_j$ implies $|\tveps_j| > \ttil_j$ for $\beta_j=0$.  
Since $|\hbeta_j-\tbeta_j|+|\that_j-\ttil_j|\le 2{c}_n\ttil_j$, 
$|\hbeta_j|\le \that_j$ implies $|\tveps_j| > \ttil_j$ for $|\beta_j|>(2+2{c}_n)\ttil_j$. Thus, 
\bes
P\Big(\{j: |\beta_j|> (2+2{c}_n)\ttil_j\} \subseteq \Shat^{(thr)} \subseteq \{j:\beta_j\neq 0\}\Big)
\ge P\{\Omega_n^c\}+p P\{|\tveps_j| > \ttil_j\}. 
\ees
Hence, (\ref{cor-2-2}) follows from $P\{|\tveps_j| > \ttil_j\}=2\Phi(-L_0)\le \alpha/p$. $\hfill\square$ 
 
\medskip
{\bf Proof of Theorem \ref{th-2}.} Due to the scale invariance of (\ref{scaled-Lasso}) and (\ref{lse-after}), 
we assume $\sigma=1$ without loss of generality. 
Let $\bh=\hbbeta^{(init)}-\bbeta$ and $z^*=\|\bX^T\bep/n\|_\infty/\sigma^*$. 
By (\ref{sparse-beta}), we have $\|\bbeta_{S^c}\|_1\le\lam_{univ}s$ and $|S|\le s$. 
Let $\{\mu_*,C_1,C_2\}$ be as in (\ref{remark-th-2-1}) and define  
\bes
\xi' = (1-\nu_0)(\xi+1)-1,\ 
\tau_*^2 = (\lam_0/\sigma^*)(\xi'+1)
\max\Big\{\frac{\lam_{univ}s}{\nu_0},\frac{\sigma^*\lam_0 s}{2(1-\nu_0)\kappa^2(\xi,S)}\Big\}. 
\ees
In the event $z^*\le (1-\tau_*^2)\lam_0(\xi'-1)/(\xi'+1)$, Theorem 2 of \cite{SunZ11} gives 
\bel{pf-th-2-1}
\max\{1-\hsigma/\sigma^*,1-\sigma^*/\hsigma\} \le \tau_*^2,\ 
\|\bh\|_1 \le (\sigma^*/\lam_0)\tau_*^2/(1-\tau_*^2) 
\eel
due to $\xi = (\xi'+\nu_0)/(1-\nu_0)$. 
Since $\kappa^2(\xi,S)\ge {c}_0$, in the event $\sigma^*>1/2$, 
\bes
\frac{\tau_*^2}{\xi+1} 
\le \max\Big\{\frac{2\lam_0\lam_{univ}s}{\nu_0/(1-\nu_0)},\frac{\lam_0^2 s}{2{c}_0}\Big\}
\le \max\Big\{\frac{(2/A)\lam_0^2s}{\nu_0/(1-\nu_0)},\frac{\lam_0^2s}{2{c}_0}\Big\}
= \frac{\tau_0^2\lam_0^2s}{A^2(1+\xi)\mu_*}.
\ees
Since $(2s/n)\log(p/\eps)\le \mu_*$, this gives $\tau_*^2 \le (\tau_0^2/\mu_*)(2s/n)\log(p/\eps)
=C_2(2s/n)\log(p/\eps)\le\tau_0^2$ in (\ref{pf-th-2-1}). 
In addition, (\ref{pf-th-2-1}) gives 
\bes
\|\bh\|_1\le \frac{\sigma^*\tau_*^2}{\lam_0(1-\tau_*^2)}
\le \frac{\sigma^*C_2s\lam_0^2}{A^2\lam_0(1-\tau_0^2)}
\le \sigma^*C_1s\sqrt{(2/n)\log(p/\eps)}. 
\ees
Thus, the union of the events in (\ref{ell_1-err-bd}) and (\ref{sigma-err-bd}) has at most probability 
\bes
p_n = P\big\{z^*\ge (1-\tau_*^2)\lam_0(\xi'-1)/(\xi'+1) \hbox{ or } \sigma^*<1/2\big\}. 
\ees

We prove below $p_n\le\eps$. 
Since $\xi'+1 = (1-\nu_0)(\xi+1)$, we have 
\bes
(1-\tau_*^2)\lam_0(\xi'-1)/(\xi'+1) \ge (1-\tau_0^2)\lam_0\{\xi - (1+\nu_0)/(1-\nu_0)\}/(\xi+1)
=\lam_0/A. 
\ees
Thus, with $p_n'=P\{\sigma^*<1/2\} = P\{\chi_n^2< n/4\}$, Theorem 2 of \cite{SunZ11} gives 
\bes
p_n\le  (1+\eps_{n-1})\eps/\{\pi\log(p/\eps)\}^{1/2} + p_n'
\ees
with $\eps_m=\{2/(m-1)\}^{1/2}\Gamma((m+1)/2)/\Gamma(m/2)$. 
Since $p_n' = \int_0^{n/4} t^{n/2-1}e^{-t/2}dt/\{2^{n/2}\Gamma(n/2)\}$
and $(n/2)\log(4/e)\ge \log(p/\eps)$, the Stirling formula gives 
\bes
p_n'\le \frac{(n/8)^{n/2}}{\Gamma(n/2+1)}\le \frac{(n/8)^{n/2}}{e^{-n/2}(n/2)^{n/2}\sqrt{2\pi}}
\le (e/4)^{n/2}/\sqrt{2\pi}\le \eps/(p\sqrt{2\pi}). 
\ees
Since $\eps_{n-1}\le \sqrt{\pi/2}$ for $n\ge 3$, 
$p_n \le \eps (1+\sqrt{\pi/2})/\sqrt{\pi\log p}+\eps/(p\sqrt{2\pi}) \le \eps$ for $p\ge 7$. 
This proves Theorem \ref{th-2} (i) for the $\{\mu^*,C_1,C_2\}$ in (\ref{remark-th-2-1}). 
The proof of Theorem \ref{th-2} (ii) follows from Theorem 3 of \cite{SunZ11} in the same way 
with somewhat different constants. We omit the details. $\hfill\square$ 

\medskip
{\bf Proof of Proposition 2.} For any $\bu\in\scrC(\xi,S)$, 
\bes
\frac{\|\bX\bu\|_2^2|S|}{n \|\bu_S\|_1^2} \ge \frac{\bu^T\hbSigma\bu}{\|\bu\|_2^2} - 
\frac{\bu^T(\bX^T\bX/n-\hbSigma)\bu}{\|\bu_S\|_1^2/|S|} 
\ge {c}_* - \frac{\lam_1\|\bu\|_1^2}{\|\bu_S\|_1^2/|S|} 
\ees
which is no smaller than ${c}_* - |S|\lam_1(1+\xi)^2\ge {c}_*/2$. 

Now assume the additional condition that $s\lam_1(1+K) \le {c}_*/2$. 
Consider $|S|\ge 1$ since the case of empty $S$ is trivial. 
Since $s\lam_1(1+K)+\lam_1 \le {c}_*/2$, we have 
$\phi_-(m,S) \ge \phi_{\min}(\hbSigma)-\lam_1(m+S)\ge {c}_* - \{|S|\lam_1(1+K)+\lam_1\}\ge {c}_*/2$.  
Similarly, $\phi_+(m,S) \le {c}_* + {c}_*/2$.  
Thus, $\phi_+(m,S)\xi^2/\kappa^2(\xi,S)\le \xi^2(c^*+{c}_*/2)/({c}_*/2)=K$. $\hfill\square$ 

\medskip
{\bf Proof of Theorem \ref{th-3}.} We first prove the bounds for $\kappa(\xi,S)$ and 
$\xi^2\phi_+(m,S)/\kappa^2(\xi,S)$ in Remark \ref{remark-th3}. 
Let $\{\delta_2,\delta_0,\delta_1, \scrX'_{n,p}, K,k,\ell\}$ 
be as in Remark \ref{remark-th3}. 
Suppose $\scrX'_{n,p}$ happens. 
Let $S$ be a subset of $\{1,\ldots,p\}$ with $|S|=k$, 
$\bu$ a vector in $\scrC(\xi,S)$ with $\|\bu_S\|_1=1$, 
$A$ the union of $S$ and the set of the indices of the $\ell$ largest $|u_j|$ with $j\not\in S$, 
and $\bw$ a unit vector in $\R^n$ with $\bw^T\bX_A\bu_A=\|\bX_A\bu_A\|_2$. 
We pick a $\bu$ satisfying 
\bes
\kappa(\xi,S) = (k/n)^{1/2}\|\bX\bu\|_2 \ge (k/n)^{1/2}\bw^T\bX\bu
= (k/n)^{1/2}\big(\|\bX_A\bu_A\|_2 + \bw^T\bX_{A^c}\bu_{A^c}\big). 
\ees
Let $u_*=\|\bu_{A^c}\|_\infty$. 
Since $\bu\in \scrC(\xi,S)$, $\|\bu_{A^c}\|_1\le \|\bu_{S^c}\|_1 - \ell u_* \le \xi - \ell u_*$. 
Let $\bu_{A^c}$ be the minimizer of $\bw^T\bX_{A^c}\bu_{A^c}$ subject to 
$\|\bu_{A^c}\|_1\le\xi-\ell u_*$ and $B_0=\{j\not\in A: u_j\neq 0\}$. 
Then, $B_0$ is the index set of certain $|B_0|$ largest $|\bw^T\bx_j|$ with $j\in A^c$ 
and $|u_j|=u_*$ for $j\in B_0$ with one possible exception. 
Since $1=\|\bu_S\|_1\le k^{1/2}\|\bu_A\|_2$ and $\|\bw\|_2=1$, 
(\ref{sparse-eigen}) gives $(k/n)^{1/2}\|\bX_A\bu_A\|_2\ge \sqrt{{c}_*(1-\delta_1)}$ and 
$\|\bw^T\bX_B/n^{1/2}\|_2\le \sqrt{c^*(1+\delta_1)}$ for all $B\subseteq B_0$ with $|B|\le 4\ell$. 
For $|B_0|\ge 4\ell$, let $B_1$ be the index set of certain $4\ell$ largest $|\bw^T\bx_j|$ with $j\in B_0$, 
so that $|\bw^T\bx_j|^2/n \le c^*(1+\delta_1)/(4\ell)$ for $j\in B_0\setminus  B_1$. 
For $|B_0|\le 4\ell$, $(\bw^T\bX_{A^c}\bu_{A^c})^2/n\le c^*(1+\delta_1)\|\bu_{A^c}\|_2^2
\le c^*(1+\delta_1)u_*(\xi-u_*\ell)\le c^*(1+\delta_1)\xi^2/(4\ell)$. 
For $|B_0|> 4\ell$, $|\bw^T\bX_{A^c}\bu_{A^c}/n^{1/2}| 
\le \sqrt{c^*(1+\delta_1)u_*^24\ell}+\|\bu_{B_0\setminus B_1}\|_1\sqrt{c^*(1+\delta_1)/(4\ell)}
=\sqrt{c^*(1+\delta_1)/(4\ell)}\|\bu_{A^c}\|_1$. 
In either cases, 
\bes
\kappa(\xi,S)
&\ge& \{{c}_*(1-\delta_1)\}^{1/2} - \{kc^*(1+\delta_1)/(4\ell)\}^{1/2}\xi
\cr &=& \{{c}_*(1-\delta_1)\}^{1/2}\big(1 - \{kK/(16\ell)\}^{1/2}\big)
\ge \{{c}_*(1-\delta_1)\}^{1/2}/2. 
\ees 
Since $m-1<K|S|\le m$ implies $m\le 4\ell$, we also have $\xi^2\phi_+(m,S)/\kappa^2(\xi,S)\le K$. 
Thus, the conditions on $\kappa(\xi,S)$ and $\phi_\pm(m,S)$ of $\scrX_{s,n,p}({c}_*,\delta_1,\xi,K)$ 
hold in $\scrX'_{n,p}$. 

By Proposition \ref{prop-1} (ii), 
the conditions on $\eta_j$ and $\tau_j$ of $\scrX_{s,n,p}({c}_*,\delta_1,\xi,K)$ hold when 
\bel{pf-th-3-1}
\min_{j\le p}\hsigma_j^2(\lam_0)/\sigma_j^2 \ge (1+\kappa_0)^2/2,\ \lam_0
=(1+\kappa_1)^{-1}3\sqrt{(\log p)/n}.
\eel
Let $\tsigma_j(\lam)$ be the scaled Lasso estimator of the noise level in the regression model 
\bel{pf-th-3-2}
\tbx_j = \sum_{k\neq j} \tgamma_{jk}\bx_k + \bep_j,\ 
\tgamma_{jk}= - \sigma_j^2\Theta_{jk}\|\tbx_k\|_2/\sqrt{n}. 
\eel
Since the scaled Lasso is scale invariant, 
$\hsigma_j(\lam_0)=\tsigma(\lam_0)\sqrt{n}/\|\tbx_j\|_2$ by (\ref{LM-j}). 
Since ${c}_*\le \sigma_j^2=1/\Theta_{jj} \le 1$ by (\ref{Theta}), 
(\ref{pf-th-3-1}) is a question about the consistency of the scaled Lasso estimator 
$\tsigma(\lam_0)$ in the regression model (\ref{pf-th-3-2}). 

Let $\delta_3\in (0,1)$ with $(1-\delta_3)(1+\delta_3)^{-1}=(1+\kappa_0)^{1/2}2^{-1/4}$ and 
\bes
\scrX_{n,p}''=\{ \max_j|1-\|\tbx_j\|_2/\sqrt{n}|\le\delta_3, 
\max_j|1-\|\bep_j\|_2/(\sigma_j\sqrt{n})|\le\delta_3\}\cap\scrX_{n,p}'.
\ees 
We have $P\{\scrX_{n,p}''\}\ge 1 -  2e^{-n\delta_2}$, taking a smaller $\delta_2$ if necessary. 
Consider the event $\scrX_{n,p}''$. 
Since $(\Theta_{jk}, k\neq j)^T\in\scrB_1(s,\lam_{univ})$ for all $j$, 
the coefficients $\tgamma_{jk}$ in (\ref{pf-th-3-2}) satisfy 
\bes
\sum_{k\neq j}\min\{|\tgamma_{jk}|/(\sigma_j\lam_{univ}),1\}
\le \sum_k\min\{(1+\delta_3)|\Theta_{jk}|/\lam_{univ}, 1\} \le (1+\delta_3)s. 
\ees
We treat $\lam_0$ as $(1+\kappa_1)^{-1}3\sqrt{(\log p)/n}=A\sqrt{(2/n)\log(p^4)}$ with 
$A = (1+\kappa_1)^{-1}3/\sqrt{8}>1$. 
By checking regularity conditions as in Remarks \ref{remark-th2} and \ref{remark-th3}, 
the scaled Lasso error bound for noise estimation gives 
\bes 
P\big\{ \tsigma_j(\lam_0)\sqrt{n}/\|\bep_j\|_2 \ge (1-\delta_3)/(1+\delta_3), \scrX_{n,p}'' \big\} \le 1/p^3. 
\ees
In the same event, $\hsigma_j(\lam_0)/\sigma_j = (\tsigma_j(\lam_0)/\sigma_j)\sqrt{n}/\|\tbx_j\|_2 
\ge (\|\bep_j\|_2/\sigma)\|\tbx_j\|_2^{-1}(1-\delta_3)/(1+\delta_3)
\ge (1-\delta_3)^2/(1+\delta_3)^2=(1+\kappa_0)/\sqrt{2}$. 
This gives (\ref{pf-th-3-1}) in the intersection of these events  
and completes the proof. $\hfill\square$



\bibliographystyle{amsalpha}
\bibliography{CI-121014}

\providecommand{\bysame}{\leavevmode\hbox to3em{\hrulefill}\thinspace}
\providecommand{\MR}{\relax\ifhmode\unskip\space\fi MR }
\providecommand{\MRhref}[2]{%
  \href{http://www.ams.org/mathscinet-getitem?mr=#1}{#2}
}
\providecommand{\href}[2]{#2}
\begin{thebibliography}{SBvdG10}

\bibitem[Ant10]{Antoniadis10}
A.~Antoniadis, \emph{Comments on: $\ell_1$-penalization for mixture regression
  models}, Test \textbf{19} (2010), no.~2, 257--258.

\bibitem[BBZ10]{BerkBZ10}
R.~Berk, L.B. Brown, and L.~Zhao, \emph{Statistical inference after model
  selection}, Journal of Quantitative Criminology \textbf{26} (2010), 217--236.

\bibitem[BCW11]{BelloniCW11}
Alexandre Belloni, Victor Chernozhukov, and Lie Wang, \emph{Square-root lasso:
  Pivotal recovery of sparse signals via conic programming}, Biometrika
  \textbf{98} (2011), no.~4, 791--806.

\bibitem[BL08]{BickelL08a}
Peter~J. Bickel and Elizaveta Levina, \emph{Regularized estimation of large
  covariance matrices}, Annals of Statistics \textbf{36} (2008), no.~1,
  199--227.

\bibitem[BRT09]{BickelRT09}
Peter Bickel, Yaacov Ritov, and Alexandre Tsybakov, \emph{Simultaneous analysis
  of {L}asso and {D}antzig selector}, Annals of Statistics \textbf{37} (2009),
  no.~4, 1705--1732.

\bibitem[BvdG11]{BuhlmannGeer11}
Peter B{\"u}hlmann and Sara van~de Geer, \emph{Statistics for high-dimensional
  data: Methods, theory and applications}, Springer, New York, 2011.

\bibitem[CDS01]{ChenDS01}
Scott~Shaobing Chen, David~L. Donoho, and Michael~A. Saunders, \emph{Atomic
  decomposition by basis pursuit}, SIAM Review \textbf{43} (2001), 129--159.

\bibitem[CT05]{CandesT05}
Emmanuel~J. Candes and Terence Tao, \emph{Decoding by linear programming}, IEEE
  Trans. on Information Theory \textbf{51} (2005), 4203--4215.

\bibitem[CT07]{CandesT07}
E.~Candes and T.~Tao, \emph{The dantzig selector: statistical estimation when
  $p$ is much larger than $n$ (with discussion)}, Annals of Statistics
  \textbf{35} (2007), 2313--2404.

\bibitem[DJ94]{DonohoJ94}
D.~L. Donoho and I.~Johnstone, \emph{Minimax risk over $\ell_p$--balls for
  $\ell_q$--error}, Probability Theory and Related Fields \textbf{99} (1994),
  277--303.

\bibitem[DS01]{DavidsonS01}
K.~Davidson and S.~Szarek, \emph{Local operator theory, random matrices and
  banach spaces}, Handbook on the Geometry of Banach Spaces, vol.~1, 2001.

\bibitem[FF93]{FrankF93}
I.E. Frank and J.H. Friedman, \emph{A statistical view of some chemometrics
  regression tools (with discussion)}, Technometrics \textbf{35} (1993),
  109--148.

\bibitem[FL01]{FanL01}
Jianqing Fan and Runze Li, \emph{Variable selection via nonconcave penalized
  likelihood and its oracle properties}, Journal of the American Statistical
  Association \textbf{96} (2001), 1348--1360.

\bibitem[FL08]{FanL08}
Jianqing Fan and Jinchi Lv, \emph{Sure independence screening for ultrahigh
  dimensional feature space (with discussion)}, J. R. Statist. Soc. \textbf{B,
  70} (2008), 849--911.

\bibitem[FL10]{FanL10}
\bysame, \emph{A selective overview of variable selection in high dimensional
  feature space}, Statistica Sinica \textbf{20} (2010), 101--148.

\bibitem[FP04]{FanP04}
J.~Fan and H.~Peng, \emph{On non-concave penalized likelihood with diverging
  number of parameters}, Annals of Statistics \textbf{32} (2004), 928--961.

\bibitem[GR04]{GreenshteinR04}
E.~Greenshtein and Y.~Ritov, \emph{Persistence in high--dimensional linear
  predictor selection and the virtue of overparametrization}, Bernoulli
  \textbf{10} (2004), 971--988.

\bibitem[Gre06]{Greenshtein06}
E.~Greenshtein, \emph{Best subset selection, persistence in high-dimensional
  statistical learning and optimization under $\ell_1$ constraint}, Annals of
  Statistics \textbf{34} (2006), 2367--2386.

\bibitem[HMZ08]{HuangMZ08}
J.~Huang, S.~Ma, and C.-H. Zhang, \emph{Adaptive lasso for sparse
  high-dimensional regression models}, Statistica Sinica \textbf{18} (2008),
  1603--1618.

\bibitem[HZ12]{HuangZ12}
Jian Huang and Cun-Hui Zhang, \emph{Estimation and selection via absolute
  penalized convex minimization and its multistage adaptive applications},
  Journal of Machine Learning Research \textbf{13} (2012), 1809--1834.

\bibitem[Joh98]{Johnstone98}
Iain Johnstone, \emph{Gaussian estimation: Sequence and wavelet models}, 1998.

\bibitem[KCO08]{KiChOh08}
Yongdai Kim, Hosik Choi, and Hee-Seok Oh, \emph{Smoothly clipped absolute
  deviation on high dimensions}, Journal of American Statistical Association
  \textbf{103} (2008), 1665--1673.

\bibitem[KLT11]{KoltchinskiiLT11}
V.~Koltchinskii, K.~Lounici, and A.~B. Tsybakov, \emph{Nuclear-norm
  penalization and optimal rates for noisy low-rank matrix completion}, The
  Annals of Statistics \textbf{39} (2011), 2302--2329.

\bibitem[Kol09]{Koltchinskii09}
V.~Koltchinskii, \emph{The dantzig selector and sparsity oracle inequalities},
  Bernoulli \textbf{15} (2009), 799--828.

\bibitem[LM11]{LaberM11}
E.~Laber and S.A. Murphy, \emph{Adaptive confidence intervals for the test
  error in classification (with discussion)}, Journal of the American
  Statistical Association \textbf{106} (2011), 904--913.

\bibitem[LP06]{LeebP06}
Hannes Leeb and Benedikt~M. Potscher, \emph{Can one estimate the conditional
  distribution of post-model-selection estimators?}, The Annals of Statistics
  \textbf{34} (2006), 2554--2591.

\bibitem[MB06]{MeinshausenB06}
Nicolai Meinshausen and Peter B{\"u}hlmann, \emph{High-dimensional graphs and
  variable selection with the lasso}, Annals of Statistics \textbf{34} (2006),
  1436--1462.

\bibitem[MB10]{MeinshausenB10}
\bysame, \emph{Stability selection (with discussion)}, Journal of the Royal
  Statistical Society, B \textbf{72} (2010), 417--473.

\bibitem[MY09]{MeinshausenY09}
N.~Meinshausen and B.~Yu, \emph{Lasso-type recovery of sparse representations
  for high-dimensional data}, Annals of Statistics \textbf{37} (2009),
  246--270.

\bibitem[SBvdG10]{StadlerBG10}
N.~St\"{a}dler, P.~B\"{u}hlmann, and S.~van~de Geer,
  \emph{$\ell_1$-penalization for mixture regression models (with discussion)},
  Test \textbf{19} (2010), no.~2, 209--285.

\bibitem[SZ10]{SunZ10}
Tingni Sun and Cun-Hui Zhang, \emph{Comments on: $\ell_1$-penalization for
  mixture regression models}, Test \textbf{19} (2010), no.~2, 270--275.

\bibitem[SZ11]{SunZ11}
Tungni Sun and Cun-Hui Zhang, \emph{Scaled sparse linear regression}, Tech.
  Report arXiv:1104.4595, arXiv, 2011.

\bibitem[Tib96]{Tibshirani96}
R.~Tibshirani, \emph{Regression shrinkage and selection via the lasso}, Journal
  of the Royal Statistical Society: Series B (Statistical Methodology)
  \textbf{58} (1996), 267--288.

\bibitem[Tro06]{Tropp06}
J.~A. Tropp, \emph{Just relax: convex programming methods for identifying
  sparse signals in noise}, IEEE Transactions on Information Theory \textbf{52}
  (2006), 1030--1051.

\bibitem[vdGB09]{vandeGeerB09}
S.~van~de Geer and P.~B{\"u}hlmann, \emph{On the conditions used to prove
  oracle results for the lasso}, Electronic Journal of Statistics \textbf{3}
  (2009), 1360--1392.

\bibitem[Wai09a]{Wainwright09b}
M.~J. Wainwright, \emph{Information-theoretic limitations on sparsity recovery
  in the high-dimensional and noisy setting}, IEEE Transactions on Information
  Theory \textbf{55} (2009), 5728--5741.

\bibitem[Wai09b]{Wainwright09}
\bysame, \emph{Sharp thresholds for noisy and high--dimensional recovery of
  sparsity using $\ell_1$--constrained quadratic programming (lasso)}, IEEE
  Transactions on Information Theory \textbf{55} (2009), 2183--2202.

\bibitem[YZ10]{YeZ10}
Fei Ye and Cun-Hui Zhang, \emph{Rate minimaxity of the lasso and dantzig
  selector for the $\ell_q$ loss in $\ell_r$ balls}, Journal of Machine
  Learning Research \textbf{11} (2010), 3481--3502.

\bibitem[ZH08]{ZhangH08}
Cun-Hui Zhang and Jian Huang, \emph{The sparsity and bias of the {L}asso
  selection in high-dimensional linear regression}, Annals of Statistics
  \textbf{36} (2008), no.~4, 1567--1594.

\bibitem[Zha05]{Zhang05}
Cun-Hui Zhang, \emph{General empirical bayes wavelet methods and exactly
  adaptive minimax estimation1}, The Annals of Statistics \textbf{33} (2005),
  54--100.

\bibitem[Zha09]{Zhang09-l1}
Tong Zhang, \emph{Some sharp performance bounds for least squares regression
  with {$L_1$} regularization}, Ann. Statist. \textbf{37} (2009), no.~5A,
  2109--2144.

\bibitem[Zha10]{Zhang10-mc+}
Cun-Hui Zhang, \emph{Nearly unbiased variable selection under minimax concave
  penalty}, The Annals of Statistics \textbf{38} (2010), 894--942.

\bibitem[Zha11a]{Zhang11-foba}
Tong Zhang, \emph{Adaptive forward-backward greedy algorithm for learning
  sparse representations}, IEEE Transactions on Information Theory \textbf{57}
  (2011), 4689--4708.

\bibitem[Zha11b]{Zhang11-multistage}
\bysame, \emph{Multi-stage convex relaxation for feature selection}, Tech.
  Report arXiv:1106.0565, arXiv, 2011.

\bibitem[ZL08]{ZouL08}
Hui Zou and Runze Li, \emph{One-step sparse estimates in nonconcave penalized
  likelihood models}, Annals of Statistics \textbf{36} (2008), no.~4,
  1509--1533.

\bibitem[Zou06]{Zou06}
Hui Zou, \emph{The adaptive lasso and its oracle properties}, Journal of the
  American Statistical Association \textbf{101} (2006), 1418--1429.

\bibitem[ZY06]{ZhaoY06}
Peng Zhao and Bin Yu, \emph{On model selection consistency of {L}asso}, Journal
  of Machine Learning Research \textbf{7} (2006), 2541--2567.

\bibitem[ZZ11]{ZhangZ11}
Cun-Hui Zhang and Tong Zhang, \emph{A general theory of concave regularization
  for high dimensional sparse estimation problems}, Tech. Report
  arXiv:1108.4988, arXiv, 2011.

\end{thebibliography}
\end{document}